\documentclass[twocolumn]{aastex62}

\usepackage{graphicx}	
\usepackage{amsmath}	
\usepackage{amssymb}	
\usepackage{bm}
\usepackage{float}
\usepackage{hyperref}
\usepackage{algorithm}
\usepackage[noend]{algpseudocode}
\usepackage[FIGTOPCAP]{subfigure}
\usepackage{tabularx}
\usepackage{booktabs}
\usepackage{threeparttable}

\allowdisplaybreaks

\def\ln{\mathrm{ln}}
\def\log10{\mathrm{log}_{10}}

\def\flux{\mathbf{f}}

\def\sps{\boldsymbol{\varphi}}

\def\data{\mathbf{d}}
\def\Data{\mathbf{D}}
\def\noise{\boldsymbol{\sigma}}

\def\hyper{\boldsymbol{\psi}}

\def\unc{\boldsymbol{\chi}}
\def\nuisance{\boldsymbol{\eta}}
\def\zp{\alpha_\mathrm{ZP}}
\def\em{\beta_\mathrm{EM}}

\graphicspath{{./}{figures/}}

\received{}
\revised{}
\accepted{}
\submitjournal{ApJS}

\shorttitle{Galaxy population modeling from COSMOS}
\shortauthors{Alsing et al.}

\begin{document}

\title{\texttt{pop-cosmos}: A comprehensive picture of the galaxy population from COSMOS data}

\correspondingauthor{Justin Alsing}
\email{justin.alsing@fysik.su.se}

\author[0000-0003-4618-3546]{Justin Alsing}
\affiliation{Oskar Klein Centre for Cosmoparticle Physics, Department of Physics, Stockholm University, Stockholm SE-106 91, Sweden}

\author[0009-0005-6323-0457]{Stephen Thorp}
\affiliation{Oskar Klein Centre for Cosmoparticle Physics, Department of Physics, Stockholm University, Stockholm SE-106 91, Sweden}

\author[0000-0003-1943-723X]{Sinan Deger}
\affiliation{Institute of Astronomy and Kavli Institute for Cosmology, University of Cambridge, Madingley Road, Cambridge CB3 0HA, UK}

\author[0000-0002-2519-584X]{Hiranya V. Peiris}
\affiliation{Institute of Astronomy and Kavli Institute for Cosmology, University of Cambridge, Madingley Road, Cambridge CB3 0HA, UK}
\affiliation{Oskar Klein Centre for Cosmoparticle Physics, Department of Physics, Stockholm University, Stockholm SE-106 91, Sweden}

\author[0000-0002-3962-9274]{Boris Leistedt}
\affiliation{Department of Physics, Imperial College London, Blackett Laboratory, Prince Consort
Road, London SW7 2AZ, UK}

\author[0000-0002-0041-3783]{Daniel Mortlock}
\affiliation{Department of Physics, Imperial College London, Blackett Laboratory, Prince Consort
Road, London SW7 2AZ, UK}
\affiliation{Department of Mathematics, Imperial College London, London SW7 2AZ, UK}
\affiliation{Oskar Klein Centre for Cosmoparticle Physics, Department of Physics, Stockholm University, Stockholm SE-106 91, Sweden}

\author[0000-0001-6755-1315]{Joel Leja}
\affil{Department of Astronomy \& Astrophysics, The Pennsylvania State University, University Park, PA 16802, USA}
\affil{Institute for Computational \& Data Sciences, The Pennsylvania State University, University Park, PA, USA}
\affil{Institute for Gravitation and the Cosmos, The Pennsylvania State University, University Park, PA 16802, USA}

\begin{abstract}
We present \texttt{pop-cosmos}: a comprehensive model characterizing the galaxy population, calibrated to $140,938$ ($r<25$ selected) galaxies from the Cosmic Evolution Survey (COSMOS) with photometry in $26$ bands from the ultra-violet to the infra-red. We construct a detailed forward model for the COSMOS data, comprising: a population model describing the joint distribution of galaxy characteristics and its evolution (parameterized by a flexible score-based diffusion model); a state-of-the-art stellar population synthesis (SPS) model connecting galaxies' instrinsic properties to their photometry; and a data-model for the observation, calibration and selection processes. By minimizing the optimal transport distance between synthetic and real data we are able to jointly fit the population- and data-models, leading to robustly calibrated population-level inferences that account for parameter degeneracies, photometric noise and calibration, and selection. We present a number of key predictions from our model of interest for cosmology and galaxy evolution, including the mass function and redshift distribution; the mass-metallicity-redshift and fundamental metallicity relations; the star-forming sequence; the relation between dust attenuation and stellar mass, star formation rate and attenuation-law index; and the relation between gas-ionization and star formation. Our model encodes a comprehensive picture of galaxy evolution that faithfully predicts galaxy colors across a broad redshift ($z<4$) and wavelength range.
\end{abstract}

\keywords{galaxy evolution - galaxy surveys - photometric redshifts}

\section{Introduction}
As galaxies evolve, their macroscopic (astro)physical characteristics -- stellar mass, metallicity, dust, gas and active galactic nuclei (AGN) content -- will evolve accordingly. While the detailed physics of galaxy-evolution and merger histories is not directly observable for individual galaxies, these processes determine the joint distribution of physical characteristics in the galaxy population, and how that distribution evolves over cosmic time. Constraining the joint distribution of galaxy properties in the Universe is therefore one of the main ways we can learn about galaxy evolution (see e.g. \citealp{madaudickinson2014} for a broad review).

As well as enabling galaxy evolution science, detailed characterization of the galaxy demographics over cosmic history is critical for cosmological probes that rely on observations of galaxies. Large-scale galaxy imaging surveys, which probe cosmological structure formation via galaxy clustering and weak gravitational lensing, require accurate determination of galaxy redshifts from their broad-band photometry. The physical characteristics of galaxies uniquely determine their spectral energy distributions \citep[SEDs; e.g.][]{conroy2013}, providing the link between prediction and observation in inferring redshifts from photometric data. The joint distribution of galaxy properties implicitly provides the prior over galaxy SEDs, which is critical for accurate photometric redshift estimation (especially from broad-band data: \citealp{arnouts1999, benitez2000, ilbert2006, brammer2008, tanaka2015}). In \cite{alsing2023} we recently showed that the redshift distributions of ensembles of galaxies in photometric surveys can be accurately derived via forward modeling, i.e., explicit modeling of the galaxy population, observational processes, and selection, provided those elements can be modeled with sufficiently high fidelity. Accurate estimation of redshift distributions is essential for obtaining robust and accurate cosmological constraints, and currently represents one of the main systematic challenges for both current (Stage III) and imminent (Stage IV) surveys, such as the Dark Energy Survey (DES; \citealp{des}), the Kilo Degree Survey (KiDS; \citealp{kids}) and Hyper Suprime-Cam (HSC; \citealp{hsc}), the Vera C. Rubin Observatory's Legacy Survey of Space and Time (LSST; \citealp{lsst}), and {\it Euclid} \citep{euclid}.

In addition, in order to leverage the cosmological information from small-scale galaxy-clustering, we require an understanding of how galaxies with different properties trace the underlying dark matter field (ie., galaxy bias, \citealp{sheth1999large, tinker2010large}). Detailed characterization of the galaxy population is hence a key component in the exploration and exploitation of the galaxy-halo connection (see e.g. \citealp{wechsler2018connection} for a recent review).

Furthermore, for transient cosmology (e.g.\ with type Ia supernovae; SNe Ia), understanding the properties of supernova host galaxies and how they correlate with intrinsic supernovae characteristics and observables is essential for drawing robust cosmological inferences. Host galaxy information has been shown to have relevance to supernova and transient classification \citep[see e.g.][]{foley2013, gagliano2021, gagliano2023}. For SNe Ia, there are models connecting galaxy evolution to possible progenitor channels \citep[e.g.][]{scannapieco2005, mannucci2005, mannucci2006, childress2014}. There are also poorly-understood correlations between SN Ia magnitudes and the mass or star formation rate of their hosts \citep[e.g.][]{kelly2010, sullivan2010}. There is considerable current debate in the literature regarding the root causes and nature of these correlations (see, e.g., \citealp{brout2021, thorp2021, nicolas2021, thorp2022, briday2022, meldorf2023, duarte2023, grayling2024}). Resolving this question will be essential for next generation projects, and already presents a challenge to current experiments \citep[e.g.][]{vincenzi2024}

In spite of its critical role in galaxy evolution, cosmology, and other fields, studies of the joint distribution of galaxy properties have largely focused on measuring specific scaling relations between two or three properties at a time, such as the (redshift evolving) mass function \citep{marchesini2009, ilbert2013, Muzzin2013a, moustakas2013, tomczak2014galaxy, grazian2015galaxy, song2016evolution, davidzon2017cosmos2015, wright2018gama, leja2020MF}, the mass-metallicity and fundamental metallicity relations (star-formation rate vs.\ gas metallicity vs.\ mass; \citealp{tremonti2004origin, maiolino2008amaze, mannucci2009lsd, lara2010fundamental, yates2012, lara2013galaxy, andrews2013, nakajima2014, yabe2015, salim2014, salim2015, kashino2016, cresci2019, cullen2021nirvandels, curti2020, bellstedt21, sanders2021mosdef, thorne2022devils}), the connection between dust and gas properties and star-formation histories \citep{burgarella2005star, kriek2013dust, arnouts2013encoding, reddy2015mosdef,salmon2016breaking, salim2016galex, leja2017, kaasinen18, tress2018shards, salim2020, nagaraj2022}, and the star-forming sequence (star-formation rate vs.\ mass vs.\ redshift; \citealp{daddi2007, noeske2007, karim2011, rodighiero2011, whitaker2012, whitaker2014, speagle2014, renzini2015, schreiber2015, tomczak2016, leslie2020, leja2021SFS}). In the case of spectroscopic studies, these relationships have typically been measured from carefully targeted subsets of galaxies, limiting their utility in describing the galaxy population at large. For studies based on larger photometric datasets, significant parameter degeneracies and uncertainties demand a principled (Bayesian hierarchical) approach to population-level inference, which can properly account for those effects; this has so far not been achieved.

In order to be useful in a cosmological inference context -- and to provide a more complete picture of the demographics of the galaxy population in general -- it is desirable to obtain comprehensive constraints on the joint density $P(\sps)$ of galaxy characteristics $\sps$, from a large and deep sample of galaxies, with as simple selection criteria as possible\footnote{i.e., as close to a simple flux-limited sample as possible.}, and with any selection cuts properly accounted and corrected for.

\begin{figure*}
    \centering
    \includegraphics[width=\linewidth]{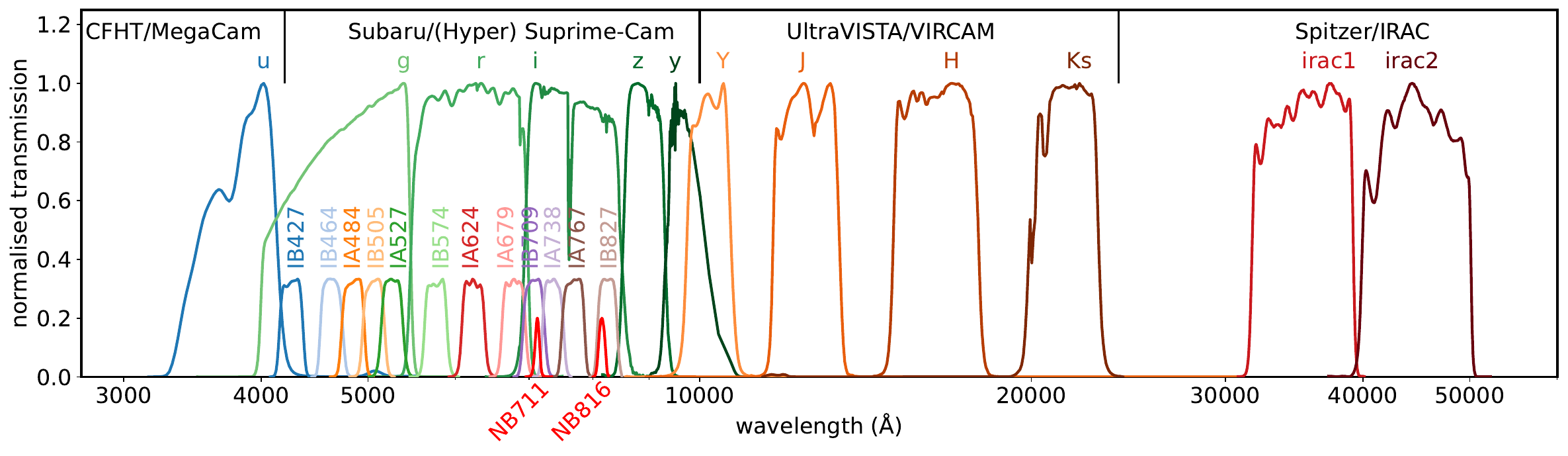}
    \caption{The subset of 26 COSMOS bands use in this work. Broadband transmission curves are rescaled to have peak transmission of 1.0, intermediate bands (``IA/B\dots'') are scaled to peak at $1/3$, and narrow bands (``NB\dots'') to peak at 0.2.}
    \label{fig:cosmos_bands}
\end{figure*}
In this work we fit a flexible, non-parametric model for the joint density of galaxy characteristics to a large, deep ($r < 25$), flux-limited sample of galaxies from the Cosmic Evolution Survey (COSMOS; \citealp{scoville2007cosmic, weaver2020}), assuming a state-of-the-art stellar population synthesis (SPS) model connecting galaxy properties with their SEDs. We construct a detailed forward model for the COSMOS data, accounting for photometric noise and calibration, selection cuts, and modeling biases at the level of the SPS predictions. We then use simulation-based inference (SBI) to jointly fit the population model along with photometric noise and calibration parameters (and modeling errors), while properly accounting and correcting for selection. The result is a robustly calibrated galaxy population model that characterizes the complex web of dependencies between galaxy characteristics, and how they evolve over cosmic time.

Our calibrated population model, which we denote \texttt{pop-cosmos}, is useful for both cosmological applications and galaxy evolution studies, and is the first joint inference of the full set of dependencies between galaxy characteristics (rather than individual scaling-relations), while properly accounting for degeneracies between parameters, calibration and selection in a principled fashion.

This represents a milestone in our ongoing effort to achieve accurate galaxy-population modeling under SPS models. In \cite{alsing2020} we developed neural emulation of SPS, achieving the ($10000\times$) speed-up required to deploy them at scale. In \cite{alsing2023} we developed the forward modeling framework, demonstrating that existing state-of-the-art population modeling can already deliver (for example) redshift distributions accurate enough for Stage III surveys. In \cite{leistedt2023hierarchical} we developed the necessary photometric- and model-calibration elements, demonstrating state-of-the-art photo-$z$ performance. Now in the present work, we combine these advances with a flexible population-model parameterization and simulation-based inference to deliver comprehensive constraints on the galaxy population from a large, deep galaxy sample with broad wavelength coverage.

This paper is structured as follows. In \S \ref{sec:data} we describe the COSMOS galaxy sample. In \S \ref{sec:model} we describe the forward model, comprising the population model (\S \ref{sec:pop-model}), SPS model (\S \ref{sec:sps-model}), calibration and noise model (\S \ref{sec:uncertainty-model}), and selection. The optimal-transport based simulation-based inference technique is described in \S \ref{sec:inference}. Results are presented in \S \ref{sec:results}, with discussion and conclusions in \S \ref{sec:discussion} and \S \ref{sec:conclusions}.
\section{Data}
\label{sec:data}
The Cosmic Evolution Survey (COSMOS), comprises deep imaging and photometry (in 44 bands) of $1.7$ million objects across $2\,\mathrm{deg}^2$ in the COSMOS field \citep{weaver2020}. We use profile-fitting based photometry from the \texttt{Farmer} (COSMOS2020) catalog \citep{weaver2020, weaver2023} in 26 of the available bands\footnote{Our complete band list is: $u$ from the Canada--France--Hawaii Telescope's MegaPrime/MegaCam; $g$, $r$, $i$, $z$, and $y$ from Subaru Hyper Suprime-Cam; $Y$, $J$, $H$, and $K_s$ from UltraVISTA \citep{mccracken2012ultravista}; \texttt{irac1} (Ch1) and \texttt{irac2} (Ch2) from Spitzer IRAC; and the Subaru Suprime-Cam intermediate and narrow bands (IB427, IB464, IA484, IB505, IA527, IB574, IA624, IA679, IB709, IA738, IA767, IB827, NB711, NB816).}, chosen to ensure well-calibrated photometry and relatively homogeneous depth across wavelengths (following \citealp{brammer2008, leistedt2023hierarchical}; see our Figure \ref{fig:cosmos_bands}). This selection excludes the Subaru Suprime-Cam broad bands (which are shallower than other filters at similar wavelengths), and the GALEX bands (which are shallow and also have broad PSFs; \citealp{weaver2020}). 

We apply the conservative combined mask, which retains the deepest regions with the greatest number of available bands while removing areas corrupted by bright stars and other artifacts \citep{weaver2020}. The catalog is prepared using the code released with the COSMOS2020 data\footnote{\url{https://github.com/cosmic-dawn/cosmos2020-readcat}.}, which applies the relevant flux corrections (including Galactic extinction) and unit conversions. We use the same star--galaxy separation criterion as \citet{weaver2020}, which is based on the $\chi^2$ estimated for star and galaxy templates in \texttt{LePhare} \citep{arnouts1999, ilbert2006} as well as morphology information from the COSMOS HST/ACS mosaics\footnote{Collectively, these cuts correspond to requiring $\texttt{lp\_type}=0$ in the COSMOS2020 catalog.}.

To construct a clean and complete analysis sample, we impose a hard magnitude cut in the $r$-band of $r < 25$, two magnitudes shallower than the 3$\sigma$ magnitude limit\footnote{\cite{weaver2023} show that the reliability of the photometric calibration (e.g., consistency between the \texttt{Farmer} and \texttt{Classic} catalogs) begins to degrade fainter than $i\gtrsim 25$, with color differences involving the $r$ band being below $0.05$ for $r<25$. Star-galaxy separation also degrades after $i\simeq 25$, with fainter sources typically being unresolved \citep{weaver2023}.}. This results in a flux-limited sample of $140,938$ galaxies, without significant additional selection effects.
\begin{table*}
    \centering
    \caption{Summary of key notation and SPS model parameters.}
    \label{tab:notation}
        \begin{tabular}{l l r r}
        \toprule
         Symbol & Description &  Details\\
         \midrule
         & \textit{Population-level hyperparameters} \\
         \cmidrule(lr){2-2}
         $\hyper$ & Population model hyperparameters (weights and biases of the diffusion model) & \S\ref{sec:pop-model}\\
         $\bm{\mu}(\flux)$ & Flux-dependent mean of uncertainty model & \S\ref{sec:uncertainty-model}\\
         $\bm{\Sigma}(\flux)$ & Flux-dependent std.\ dev.\ of uncertainty model & \S\ref{sec:uncertainty-model}\\
         $\unc$ & Uncertainty model hyperparameters (weights and biases of the MDN) & \S\ref{sec:uncertainty-model}\\
         $\bm{\alpha}_\mathrm{ZP}$ & Zero-point corrections ($\times26$) & Tab.\ \ref{tab:zps}, \S\ref{sec:structure}\\
         $\bm{\beta}_\mathrm{EM}$ & Emission line (fractional) corrections relative to \texttt{CLOUDY} ($\times44$) & Tab.\ \ref{tab:emlines}, \S\ref{sec:structure}\\
         $\bm{\gamma}_\mathrm{EM}$ & Fractional variance in emission line contributions ($\times44$) & Tab.\ \ref{tab:emlines}, \S\ref{sec:structure}\\
         $\nuisance$ & All ``nuicance'' parameters $\{\unc, \bm{\alpha}_\mathrm{ZP}, \bm{\beta}_\mathrm{EM}, \bm{\gamma}_\mathrm{EM}\}$ & \S\ref{sec:inference}\\
         \midrule
         & \textit{Galaxy-level quantities} \\
         \cmidrule(lr){2-2}
         $\sps$ & SPS model parameters & Tab.\ \ref{tab:notation}, \S\ref{sec:sps-model}\\
         $f_b^\mathrm{SPS}(\sps,z)$ & SPS model flux in band $b$ & Eq.\ \eqref{eq:sps-flux}, \S\ref{sec:structure}\\
         $\flux_b^\mathrm{EM}(\sps,z)$ & Vector of emission-line contributions to the flux in band $b$ & \S\ref{sec:structure}\\
         $f_b(\sps,z)$ & Total model flux in band $b$ & Eq.\ \eqref{eq:calibration}, \S\ref{sec:structure}\\
         $\flux$ & True model flux in all bands $\{f_{1:26}(\sps,z)\}$ & \S\ref{sec:structure}\\
         $\noise_\mathrm{P}$ & Photometric uncertainty & Eq.\ \eqref{eq:draw_unc}, \S\ref{sec:structure}, \S\ref{sec:uncertainty-model}\\
         $\noise_\mathrm{EM}$ & Uncertainty due to un-modeled emission line variations & Eq.\ \eqref{eq:emline-unc}, \S\ref{sec:structure}\\
         $\noise$ & Total noise standard deviation ($\noise^2=\noise^2_\mathrm{P}+\noise^2_\mathrm{EM}$) & \S\ref{sec:structure}\\
         $\data$ & Vector of noisy, calibrated model fluxes & Eq.\ \eqref{eq:data}, \S\ref{sec:structure}\\
         $\hat{\data}$ & Vector of observed fluxes &  \S\ref{sec:inference}\\
         $\bm{u},\bm{s}\sim\mathcal{N}(0,1)$ & Base normal random variates for population and uncertainty models & \S\ref{sec:uncertainty-model}, \S\ref{sec:inference}\\
         $\bm{n}\sim\mathcal{T}_2$ & Base Student's-$t$ variates for the noise model & \S\ref{sec:structure}, \S\ref{sec:inference}\\
         $\Data$ & Full sample of mock photometry (ie., a mock catalog realization), $\Data=\{\data\}_{1:N}$ & \S\ref{sec:inference}\\
         $\hat{\Data}$ & Observed catalog of COSMOS photometry, $\hat\Data=\{\hat\data\}_{1:N}$ & \S\ref{sec:inference}\\
         \midrule
         & \textit{Distributions and functions}\\
         \cmidrule(lr){2-2}
         $P(\sps, z|\hyper)$ & Population model (score-based diffusion model) & \S\ref{sec:pop-model}\\
         $P(\noise_\mathrm{P} | \flux;\unc)$ & Uncertainty model (mixture density network) & \S\ref{sec:uncertainty-model}\\
         $P(\mathbf{n})$ & Whitened noise distribution (independent Student's-$t$, 2 d.o.f.) & Eq. \eqref{eq:data}, \S\ref{sec:inference}\\
         $\mathcal{W}_2(\Data, \hat{\Data})$ & Optimal transport distance between $\Data$ and $\hat{\Data}$ & \S\ref{sec:inference}\\
         \midrule
         & \textit{SPS model parameters} & Prior Limits \\
         \cmidrule(lr){2-3}
         $z$ & Redshift & $[0.0, 4.5]$\\
         $\log10(M/M_\odot)$ & Stellar mass & $[7.0, 13.0]$\\
         $\Delta\log10(\text{SFR})$ & Logarithm of ratios of SFR between redshift bins ($\times6$) & $[-5.0,5.0]$\\
         $\tau_1$ & Optical depth of dust in birth cloud & $[0.0, 2.0]$\\
         $\tau_2$ & Optical depth of diffuse dust & $[0.0,4.0]$\\
         $n$ & Index of dust attenuation law & $[-1.0,0.4]$\\
         $\ln(f_\text{AGN})$ & Fractional contribution of AGN to luminosity & $[\ln(10^{-5}),\ln(3)]$\\
         $\ln(\tau_\text{AGN})$ & Optical depth of AGN dust torus & $[\ln(5), \ln(150)]$\\
         $\log10(Z_\text{gas}/Z_\odot)$ & Gas-phase metallicity & $[-2.0, 0.5]$\\
         $\log10(U_\text{gas})$ & Gas ionization & $[-4.0, -1.0]$\\
         $\log10(Z/Z_\odot)$ & Stellar metallicity & $[-1.98, 0.19]$\\
         \bottomrule
        \end{tabular}
\end{table*}
\section{Model}
\label{sec:model}
Our generative model for photometric galaxy survey data comprises a sequence of steps for simulating mock galaxy catalogs, which can then be compared against the observed data in a simulation-based inference setting for estimating the population-level parameters of interest (as described in \S \ref{sec:inference}).
Notation is summarized in Table \ref{tab:notation}, the overall model structure and key components are outlined in \S \ref{sec:structure}, while the detailed assumptions about each model component are given in \S \ref{sec:pop-model}--\ref{sec:uncertainty-model}. The logical flow of our forward model  is also summarized in Figure \ref{fig:schematic} (left panel).
\subsection{Generative model structure}
\label{sec:structure}
Our generative model proceeds in the following sequence of steps:
\begin{enumerate}
\item \textbf{Draw galaxy parameters}: SPS parameters $\sps$ and redshifts $z$ are drawn for each galaxy from the population-model $P(\sps, z | \hyper)$. Inference of the population model parameters $\hyper$ is the main target of our analysis. We parameterize $P(\sps, z | \hyper)$ as a score-based diffusion model (\citealp{song2019generative, song2020denoising, song2020score}; see \S \ref{sec:pop-model} for details);
\item \textbf{Compute photometry}: Rest-frame spectral energy distributions (SEDs) $l(\lambda; \sps)$ are calculated for each galaxy, given its SPS parameters $\sps$ and the assumed SPS model. The photometry $f_b$ in each band $b$, for each galaxy, is then obtained by:
\begin{align}
    f&^\mathrm{SPS}_b(\sps, z) = \nonumber \\
    & \frac{(1+z)^{-1}}{4\pi d_L^2(z)}\int_0^\infty l(\lambda/(1+z); \sps)e^{-\tau(z, \lambda)}W_b(\lambda)d\lambda,
    \label{eq:sps-flux}
\end{align}
where $d_L(z)$ the luminosity distance for redshift $z$, $\tau(z, \lambda)$ is the optical depth of the inter-galactic medium, and $W_b(\lambda)$ are the band-passes for each band $b$. We assume a state-of-the-art $16$-parameter SPS model, detailed in \S \ref{sec:sps-model};
\item \textbf{Calibrate photometry}: Measured photometry is subject to calibration biases. We apply zero-point corrections $\zp$ per band. The SPS model, too, will be subject to small biases due to approximations and missing model components. For example, emission-line predictions are often only accurate at the $\mathcal{O}(10\%)$ level or less \citep{leistedt2023hierarchical}, with variation arising from both the SPS treatment used \citep[e.g.][]{byler2017}, and the scheme used to compute line intensities (quantum mechanical vs.\ semi-classical, etc.; see \citealp{guzman2017, ferland2017}). In this step, we apply the zero-point $\zp$ and emission-line $\em$ corrections to the SPS model photometry from step 2:
\begin{align}
\label{eq:calibration}
f_{b}(\sps, z)= \alpha_\mathrm{ZP}[f^\mathrm{SPS}_{b}(\sps, z) + \bm{\beta}_\mathrm{EM}\cdot \flux^\mathrm{EM}_{b}(\sps, z)],
\end{align}
where $\flux^\mathrm{EM}_{b}(\sps, z)$ is the vector of emission-line contributions to the photometry for band $b$. We include emission-line corrections to the $44$ strongest emission-lines, following \citet{leistedt2023hierarchical} (see our Table \ref{tab:emlines} for a list of included lines);
\item \textbf{Draw uncertainties}: We draw photometric uncertainties (noise variances) for each galaxy from an uncertainty model, $\noise_\mathrm{P} \leftarrow P(\noise_\mathrm{P} | \flux ; \unc)$. The uncertainty model $P(\noise_\mathrm{P} | \flux ; \unc)$, parameterized by $\unc$, describes variation in photometric uncertainties from galaxy to galaxy due to heterogeneous observing conditions and strategy, varying difficulty in extracting photometry from galaxies with different morphologies and geometries, and the scaling of uncertainties with flux due to the Poisson photon count contribution to the measurement errors. Construction of the uncertainty model is detailed in \S \ref{sec:uncertainty-model}.

We model an additional source of photometric uncertainty arising from variability in the emission-line contributions to each galaxy (relative to \texttt{CLOUDY} predictions); these depend on the detailed micro-structure of the galaxy and are not captured in the SPS parameterization. To this end, we construct emission-line contributions to the photometric uncertainties in each band, parameterized as
\begin{align}
    \sigma_{\mathrm{EM},b} = \bm{\gamma}_\mathrm{EM} (\bm{\beta}_\mathrm{EM} + 1)\cdot \flux^\mathrm{EM}_{b},
    \label{eq:emline-unc}
\end{align}
where $\bm{\gamma}_\mathrm{EM}$ represent the (fractional) variance in the emission-line contributions for each of the $44$ lines included (Table \ref{tab:emlines}). The total photometric uncertainty for each galaxy is then given by the quadrature sum of measurement and emission-line contributions $\noise^2 = \noise_\mathrm{P}^2 + \bm{\sigma}_\mathrm{EM}^2$;
\item \textbf{Add noise}: We add noise to the calibrated model photometry from step 3, given the photometric uncertainties from step 4, assuming independent Student's-t errors on each band (with two degrees-of-freedom),
\begin{align}
    \data &= \flux + \noise\odot\mathbf{n}, \nonumber \\
    P(n) &= \frac{1}{2\sqrt{2}(1+n^2/2)^{3/2}},
    \label{eq:data}
\end{align}
where $\data$ is the vector of noisy (calibrated) fluxes, $\odot$ denotes element-wise multiplication, and $\flux$ denotes the vector of model fluxes (i.e.\ all the $f_b(\sps,z)$ computed in step 3);
\item \textbf{Apply selection}: Galaxies are selected into the sample following the same photometric cuts that were applied to the data (\S \ref{sec:data}).
\end{enumerate}
This generative process represents a complete description of our model assumptions, or equivalently, our simulation pipeline for generating mock galaxy catalog data. Simulated catalogs generated in this way can be compared to the data in a simulation-based inference setting in order to estimate the population-level parameters (see \S \ref{sec:inference}).

In the following sections, we give more details of the population model (\S \ref{sec:pop-model}), SPS model (\S \ref{sec:sps-model}), and uncertainty model (\S \ref{sec:uncertainty-model}) assumptions. The simulation-based fitting procedure is then described in \S \ref{sec:inference}.

\begin{table}
    \centering
    \caption{Inferred zero-point corrections (see Eq.\ \ref{eq:calibration}).}
    \label{tab:zps}
    \begin{tabular}{c c c c}
        \toprule
        \multicolumn{2}{c}{Broad bands} & \multicolumn{2}{c}{Narrow bands}\\
        \cmidrule(r){1-2}  \cmidrule(l){3-4}
        Band & $\alpha_\text{ZP}$ & Band & $\alpha_\text{ZP}$\\
        \midrule
         $u$ & 1.001912 & IB427 & 0.969370\\
        $g$ & 1.075040 & IB464 & 0.983500\\
        $r$ & 1.063653 & IA484 & 1.011142\\
        $i$ & 1.007897 & IB505 & 0.998376\\
        $z$ & 1.012354 & IA527 & 0.984130\\
        $y$ & 1.038180 & IB574 & 0.939463\\
        $Y$ & 1.009543 & IA624 & 1.001962\\
        $J$ & 0.996319 & IA679 & 1.139179\\
        $H$ & 0.973534 & IB709 & 0.972257\\
        $K_s$ & 1.051483 & IA738 & 0.959059\\
        \texttt{irac1} & 0.960127 & IA767 & 0.961810 \\
        \texttt{irac2} & 0.932108 & IB827 & 0.931655\\
        && NB711 & 0.976505 \\
        && NB816 & 0.936989 \\
        \bottomrule
    \end{tabular}
\end{table}

\begin{table}
    \centering
    \caption{List of emission lines used, with our inferred fractional corrections ($\beta_\mathrm{EM}$) and variances ($\gamma_\mathrm{EM}$). Line wavelengths are from \citet{byler2017}, with the list of 44 selected lines being from \citet{leistedt2023hierarchical}.}
    \label{tab:emlines}
    \begin{tabular}{l c c c}
        \toprule
        Line & $\lambda_\mathrm{EM}$ (\AA) & $\beta_\mathrm{EM}$ & $\gamma_\mathrm{EM}$\\
        \midrule
        C \textsc{ii}] 2326 & 2326.11 & $-2.202\times 10^{-5}$ & $1.425\times 10^{-13}$\\\relax 
        [O \textsc{iii}] 2321 & 2321.66 & $7.630\times 10^{-4}$ & $1.424\times 10^{-13}$\\\relax 
        [O \textsc{i}] 6302 & 6302.05 & $-7.819\times 10^{-5}$ & $1.464\times 10^{-13}$\\\relax 
        [S \textsc{ii}] 4070 & 4069.75 & $-7.173\times 10^{-3}$ & $1.433\times 10^{-13}$\\\relax 
        H \textsc{i} (Ly-$\alpha$) & 1215.67 & $-3.610\times 10^{-4}$ & $1.455\times 10^{-13}$\\\relax 
        [Al \textsc{ii}] 2670 & 2669.95 & $2.866\times 10^{-3}$ & $1.425\times 10^{-13}$\\\relax 
        [Ar \textsc{iii}] 7753 & 7753.19 & $-5.054\times 10^{-3}$ & $1.427\times 10^{-13}$\\\relax 
        H \textsc{i} (Pa-7) & 9017.80 & $-3.288\times 10^{-3}$ & $1.425\times 10^{-13}$\\\relax 
        [Al \textsc{ii}] 2660 & 2661.15 & $2.906\times 10^{-3}$ & $1.425\times 10^{-13}$\\\relax 
        [S \textsc{iii}] 6314 & 6313.81 & $-1.134\times 10^{-3}$ & $1.426\times 10^{-13}$\\\relax 
        H \textsc{i} (Pa-6) & 9232.20 & $-1.846\times 10^{-3}$ & $1.426\times 10^{-13}$\\\relax 
        [S \textsc{iii}] 3723 & 3722.75 & $-4.681\times 10^{-3}$ & $1.425\times 10^{-13}$\\\relax 
        Mg \textsc{ii} 2800 & 2803.53 & $-3.293\times 10^{-3}$ & $1.434\times 10^{-13}$\\\relax 
        H \textsc{i} (Pa-5) & 9548.80 & $-1.264\times 10^{-3}$ & $1.427\times 10^{-13}$\\\relax 
        He \textsc{i} 7065 & 7067.14 & $-2.074\times 10^{-3}$ & $1.427\times 10^{-13}$\\\relax 
        [N \textsc{ii}] 6549 & 6549.86 & $8.271\times 10^{-4}$ & $2.163\times 10^{-13}$\\\relax 
        [S \textsc{ii}] 6732 & 6732.67 & $-5.002\times 10^{-4}$ & $3.600\times 10^{-13}$\\\relax 
        C \textsc{iii}] & 1908.73 & $-1.061\times 10^{-3}$ & $1.424\times 10^{-13}$\\\relax 
        He \textsc{i} 6680 & 6679.99 & $-1.390\times 10^{-3}$ & $1.437\times 10^{-13}$\\\relax 
        Mg \textsc{ii} 2800 & 2796.35 & $-2.295\times 10^{-3}$ & $1.460\times 10^{-13}$\\\relax 
        [S \textsc{ii}] 6717 & 6718.29 & $-1.463\times 10^{-3}$ & $1.028\times 10^{-13}$\\\relax 
        [Ar \textsc{iii}] 7138 & 7137.77 & $-1.661\times 10^{-3}$ & $1.490\times 10^{-13}$\\\relax 
        [C \textsc{iii}] & 1906.68 & $-9.706\times 10^{-4}$ & $1.425\times 10^{-13}$\\\relax 
        He \textsc{i} 4472 & 4472.73 & $4.921\times 10^{-3}$ & $1.437\times 10^{-13}$\\\relax 
        [O \textsc{iii}] 4364 & 4364.44 & $4.065\times 10^{-3}$ & $1.425\times 10^{-13}$\\\relax 
        [N \textsc{ii}] 6585 & 6585.27 & $-6.022\times 10^{-1}$ & $1.000\times 10^{-13}$\\\relax 
        [S \textsc{iii}] 9071 & 9071.10 & $-1.003\times 10^{0}$ & $1.702\times 10^{-13}$\\\relax 
        H-8 3798 & 3798.99 & $-1.548\times 10^{-3}$ & $1.465\times 10^{-13}$\\\relax 
        He \textsc{i} 3889 & 3889.75 & $-5.387\times 10^{-3}$ & $1.500\times 10^{-13}$\\\relax 
        H-7 3835 & 3836.49 & $-1.911\times 10^{-3}$ & $1.485\times 10^{-13}$\\\relax 
        [Ne \textsc{iii}] 3968 & 3968.59 & $-7.520\times 10^{-3}$ & $1.448\times 10^{-13}$\\\relax 
        He \textsc{i} 5877 & 5877.25 & $3.262\times 10^{-1}$ & $1.555\times 10^{-13}$\\\relax 
        H-6 3889 & 3890.17 & $-5.784\times 10^{-3}$ & $1.536\times 10^{-13}$\\\relax 
        [S \textsc{iii}] 9533 & 9533.20 & $-1.001\times 10^{0}$ & $2.017\times 10^{-13}$\\\relax 
        H-5 3970 & 3971.20 & $-1.486\times 10^{-1}$ & $1.689\times 10^{-13}$\\\relax 
        [O \textsc{ii}] 3726 & 3727.10 & $-1.030\times 10^{-3}$ & $1.000\times 10^{-13}$\\\relax 
        H-$\delta$ 4102 & 4102.89 & $-5.205\times 10^{-1}$ & $1.912\times 10^{-13}$\\\relax 
        [O \textsc{ii}] 3729 & 3729.86 & $2.583\times 10^{-1}$ & $1.000\times 10^{-13}$\\\relax 
        [Ne \textsc{iii}] 3870 & 3869.86 & $-8.755\times 10^{-2}$ & $1.889\times 10^{-13}$\\\relax 
        H-$\gamma$ 4340 & 4341.69 & $-3.269\times 10^{-1}$ & $1.013\times 10^{-13}$\\\relax 
        [O \textsc{iii}] 4960 & 4960.30 & $-9.501\times 10^{-3}$ & $1.000\times 10^{-13}$\\\relax 
        H-$\beta$ 4861 & 4862.71 & $-5.651\times 10^{-1}$ & $1.000\times 10^{-13}$\\\relax 
        H-$\alpha$ 6563 & 6564.60 & $-3.420\times 10^{-1}$ & $1.000\times 10^{-13}$\\\relax 
        [O \textsc{iii}] 5007 & 5008.24 & $1.063\times 10^{-1}$ & $2.633\times 10^{-2}$ \\
        \bottomrule
    \end{tabular}
\end{table}

\newpage
\subsection{Population model}
\label{sec:pop-model}
The population model $P(\sps, z | \hyper)$ is the main target of our analyses. We require a flexible parameterization for this high-dimensional density, which is capable of capturing the complex web of inter-dependencies between galaxies' properties that arise from galaxy formation and evolution physics. Advances in generative machine-learning models, such as normalizing flows \citep{rippel2013high, germain2015made, dinh2016density, papamakarios2017, grathwohl2018ffjord, chen2018neural, kingma2018glow, durkan2019, papamakarios2021normalizing} and diffusion models \citep{sohl2015deep, ho2020denoising, song2019generative, song2020score, song2020denoising, song2020improved, kingma2021variational,luo2022understanding} have provided a step change in our ability to parameterize and learn complex and high-dimensional probability distributions from data.

We parameterize $P(\sps, z | \hyper)$ using a score-based diffusion model \citep{song2019generative, song2020denoising, song2020score}, where the population-model parameters $\hyper$ are the weights and biases of the score-network (outlined below). Diffusion models have been shown to be effective flexible approximators for unknown probability distributions, are relatively inexpensive to train, and scale well to high-dimensional problems, making them ideally suited to this use-case (see \citealp{luo2022understanding} for a review).

In diffusion models, as with normalizing flows, we aim to find an invertible transform that maps from some simple base density (eg., a unit normal) to our target $p(\bm{x})$, such that we can generate samples from the target by simply transforming draws from the base-density, ie.,
\begin{align}
    & \bm{x} = \bm{f}(\bm{u}),\; \bm{u}\sim \mathcal{N}(0, 1) \nonumber \\
    & p(\bm{x}) = \mathcal{N}(\bm{f}^{-1}(\bm{x}) | 0, 1) |\bm{J}(\bm{x})|,
\end{align}
where $\bm{J} = \partial\bm{f}^{-1}(\bm{x})/\partial\bm{x}$ is the Jacobian, and the transform $\bm{f}$ must be invertible. In both normalizing flows and diffusion models, the goal is to parameterize the invertible transform $\bm{f}$ by a neural network.

In a score-based diffusion model, we begin by constructing a diffusion process $\{\bm{x}(t)\}_{t=1}^{t=T}$ (indexed by a continuous time-variable $t$) such that $\bm{x}(t=0)$ is distributed according to our target distribution, and $\bm{x}(t=T)$ has a normal distribution. This diffusion process can be described by a stochastic differential equation (SDE), which maps samples from our target distribution at $t=0$ to random noise at $t=T$:
\begin{equation}
\label{eq:sde}
    d\bm{x} = \bm{f}(\bm{x}, t)dt + g(t)d\bm{w},
\end{equation}
where $\bm{w}$ is standard Brownian motion (Wiener process). In order to generate samples from our target then, we can take samples from the base normal distribution $\bm{x}(t=T)$ and reverse the process back to $t=0$. The reverse of a diffusion process defined by Equation \eqref{eq:sde} is simply another diffusion process, defined by the reverse-time SDE \citep{anderson1982reverse, song2020score}:
\begin{align}
\label{eq:sde_reverse}
    d\bm{x} = \left[\bm{f}(\bm{x}, t) + g(t)^2\nabla_{\bm{x}} p_t(\bm{x}) \right]dt + g(t)d\bar{\bm{w}},
\end{align}
where $\bar{\bm{w}}$ is reverse-time Brownian motion, $p_t(\bm{x})$ are the marginal distributions of the diffusion process defined by Equation \eqref{eq:sde}, and $dt$ is an infitesimal step backwards in time. Hence, once the score $\nabla_{\bm{x}}p_t(\bm{x})$ of the marginals of the forward diffusion process is known as a function of time, then the reverse-process in Equation \eqref{eq:sde_reverse} can be evaluated to transform samples from the base density $\bm{x}(t=T)\sim\mathcal{N}(0,1)$ to the target $\bm{x}(t=0)$. The transform from the base-density to the target is hence completely characterized by the score of the marginals: in a score-based diffusion model, the score $\bm{s}(\bm{x}, t) = \nabla_{\bm{x}} p_t(\bm{x})$ is parameterized as a (dense) neural network, and fit by denoising score-matching \citep{hyvarinen2005estimation, song2020denoising, song2020score, song2021maximum}, or otherwise.

So far, this reverse-time diffusion process provides a means to stochastically transform from the base density to the target, via Equation \eqref{eq:sde_reverse}. However, in order to be able to evaluate the Jacobian and hence log probability, we require a deterministic (invertible) transform between the base and the target. Fortunately, for any reverse-time SDE of the form given in Equation \eqref{eq:sde_reverse}, there exists a deterministic ordinary differential equation (ODE) that has the same marginal distributions as the SDE \citep{maoutsa2020interacting, song2020score}:
\begin{equation}
\label{eq:ode}
    d\bm{x} = \left[\bm{f}(\bm{x}, t) -\frac{1}{2} g(t)^2\nabla_{\bm{x}} p_t(\bm{x}) \right]dt.
\end{equation}
Integrating this ODE from $t=T$ to $t=0$ hence provides a deterministic, invertible transform from the base density $\bm{x}(t=T)$ to the target $p(\bm{x})$, which is completely characterized by the score-function $\bm{s}(\bm{x}, t) = \nabla_{\bm{x}} p_t(\bm{x})$, and whose Jacobian can be computed. Interpreting the diffusion model as an ODE transform in this way elicits an equivalence between continuous-time normalizing flows \citep{grathwohl2018ffjord, chen2018neural} and score-based diffusion models \citep{song2020score}.

 We parameterize the score $\bm{s}(\bm{x}, t)$ as a dense network with two layers of $128$ hidden units and \texttt{tanh} activation functions. We take the so-called variance-exploding SDE \citep{song2020score} to define the forward diffusion process,
\begin{align}
    d\bm{x} = g(t)d\bm{w},\; g(t)^2 = \frac{d\sigma^2(t)}{dt},\;\sigma(t) = \sigma_0(\sigma_T/ \sigma_0)^{t/T},
\end{align}
where we choose $\sigma_0=0.01$, $\sigma_T=10$ and $T=1$, and implicitly in the variance-exploding SDE the drift-term $\bm{f}(\bm{x}, t)$ is set to zero (c.f.\ Equation \ref{eq:sde}).
\subsection{Stellar population synthesis (SPS) model}
\label{sec:sps-model}
Stellar population synthesis provides the theoretical framework linking the stellar, gas and dust content of galaxies, and their SEDs (see \citealp{conroy2013} for a review). We assume a state-of-the-art $16$-parameter SPS model, based on the milestone \texttt{Prospector}-$\alpha$ model \citep{leja2017, leja2018, leja2019np, leja2019}, but including the gas-phase ionization parameter as an additional free parameter; we found that this additional parameter was required to give reasonable inferences about the gas-phase physics. For completeness, the physical assumptions and parameters are described below.

The star formation history (SFH) is modeled as piece-wise constant, with seven
bins in time \citep[see][]{leja2019np}. The first two bins are fixed at $[0, 30]$ Myr
and $[30, 100]$ Myr respectively, to capture recent star formation. The oldest bin is fixed at $[0.85, 1] t_\mathrm{age}(z)$, where $t_\mathrm{age}(z)$ is the age of the universe at the lookback time of the galaxy. The remaining four bins are equally-spaced (logarithmically) in
time between $100$ Myr and $0.85 t_\mathrm{age}(z)$. The ratios of the log star formation rate (SFR) between adjacent SFH bins are then the free model parameters describing the SFH. This flexible six-parameter model is able to capture a rich diversity of SFH phenomenology, including both smooth and bursty star-formation histories.

Dust is modeled as separate diffuse and birth cloud dust screens, where the latter only impacts stars younger than $10$ Myr \citep{charlot2000}. The optical depths $\tau_1$ (birth cloud) and $\tau_2$ (diffuse), as well as the power-law index of the \citet{calzetti2000} attenuation curves, constitute the free parameters describing the dust model. Dust emission is powered by energy-balance.

The stellar metallicity for all stars in the galaxy is assumed to take a single value, ie., the model does not explicitly account for time-varying metallicity in the stellar population. This is generally a good approximation, although some studies suggest that metallicity evolution can improve SED modeling at the level of (typically) a few percent \citep{robotham2020prospect, bellstedt2020galaxy}.

Gas is characterized by the gas-phase metallicity and ionization state (treated as separate independent variables). Nebular line and continuum emission is generated using \texttt{CLOUDY} \citep{ferland2013, ferland2017} model grids from \citet{byler2017}. We assume that the gas-phase metallicity must be greater than or equal to the stellar metallicity (since the latter captures the light-weighted average over the stellar population, which includes older stars).

Active galactic nucleus (AGN) activity is modeled as described in \citet{leja2018}, where the fraction of the bolometric luminosity from the AGN $f_\mathrm{AGN}$ and optical depth of the AGN torus $\tau_\mathrm{AGN}$ are both included as free parameters.

Together with stellar mass and redshift, this amounts to a total of $16$ parameters characterizing each galaxy. The list of parameters and their prior limits are given in Table \ref{tab:notation}.

We assume MIST stellar evolution tracks and isochrones \citep{choi2016, dotter2016}, based on MESA \citep{paxton2010, paxton2013, paxton2015}, and a \cite{chabrier2003galactic} initial mass function (IMF). We assume a solar metallicity of $Z_\odot=0.0142$.

The SPS model is implemented in the public code Flexible Stellar Population Synthesis (\texttt{FSPS}; \citealp{conroy2009, conroy2010ii, conroy2010, conroy2010fsps}),
accessed through the \texttt{python-fsps} binding \citep{foreman2014}. We then use \texttt{speculator} \citep{alsing2020} to accelerate the SPS computation, achieving a factor of $10^4$ speed-up over \texttt{FSPS} per band, while maintaining sub-percent accuracy on the predicted fluxes\footnote{We follow the same architecture and training hyper-parameter choices as for the \texttt{Prospector}-$\alpha$ model emulators constructed in \cite{alsing2020}.}.
\subsection{Uncertainty model}
\label{sec:uncertainty-model}
The uncertainty model describes the distribution of photometric measurement uncertainties in the survey, conditional on the true source flux, $P(\noise_\mathrm{P} | \flux ; \unc)$. Following \cite{alsing2023}, we model $P(\noise_\mathrm{P} | \flux ; \unc)$ as a mixture density network (MDN; \citealp{bishop2006}). Here we use an MDN with one Gaussian component, i.e., a neural network parameterizing the mean $\bm{\mu}(\flux)$ and standard deviation $\bm{\Sigma}(\flux)$ of the distribution of photometric uncertainties, conditioned on flux. From this, photometric uncertainties can be drawn for a simulated galaxy with flux $\flux$ by drawing:
\begin{align}
    \label{eq:draw_unc}
    \noise_\mathrm{P} = \bm{\mu}(\flux) + \bm{\Sigma}(\flux)\odot \bm{s},\; \bm{s}\sim \mathcal{N}(0,1).
\end{align}
We parameterize the MDN with a single dense network with two hidden layers, with $128$ units each and \texttt{tanh} activation functions. 

By keeping the uncertainty model parameters $\unc$ free in the fitting process, we are able to fully self-calibrate the uncertainty model from the data, eliminating any reliance on the (approximate) estimated flux uncertainties in the \texttt{Farmer} catalog.
\section{Inference}
\label{sec:inference}
\begin{figure*}
\centering
\includegraphics[width = \linewidth]{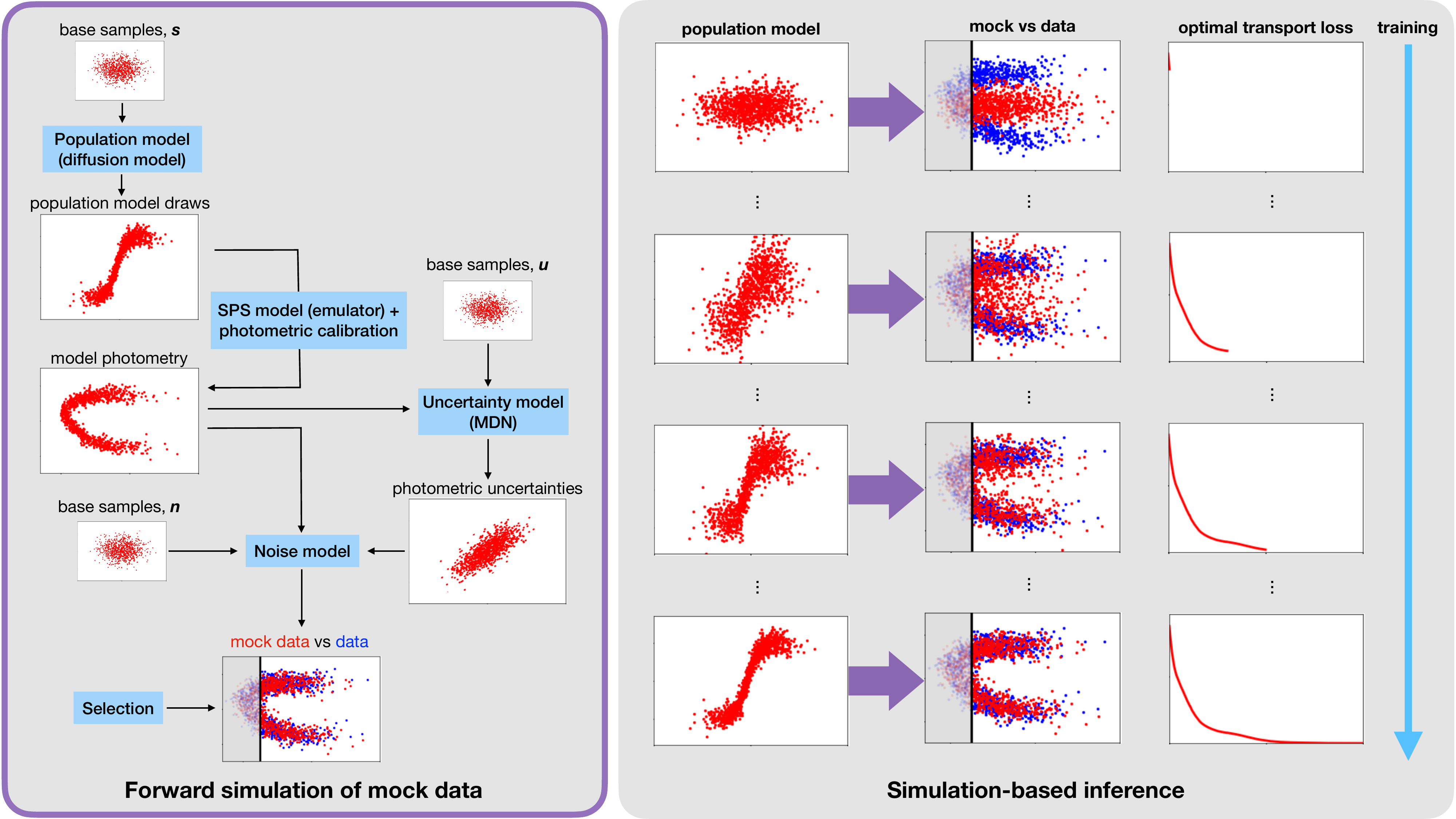}
\caption{Left: Logical flow of our forward modeling (simulation) framework described in \S \ref{sec:model}. Galaxy parameters are drawn from the population model (parameterized as a score-based diffusion model; \S~\ref{sec:pop-model}); calibrated model photometry are calculated assuming the SPS model (\S~\ref{sec:sps-model}) and calibration models (Equation \ref{eq:calibration}); photometric uncertainties are drawn from the uncertainty model (\S~\ref{sec:uncertainty-model}); photometric noise is drawn and added given the noise model (Equation \ref{eq:data}); and finally selection is applied (\S~\ref{sec:data}). Note that the three stochastic steps (the population model, uncertainty model, and noise draws) are parameterized as bijective transforms from a base density (unit normal for the population- and uncertainty-model, and Student's-t for the noise-model draws; see \S~\ref{sec:inference}). Each red and blue point represents a galaxy in the mock and real data respectively. Right: Schematic illustration of our simulation-based inference framework, optimizing the optimal transport distance between simulated (red) and real (blue) data, by gradient descent. The forward modeling stages expanded in detail on the left are represented by the purple arrows in the right-hand block. Gradients of the simulator can be obtained via automatic differentiation, by keeping the input draws from the base density fixed in the forward simulations (see \S~\ref{sec:inference}). Note that by performing inference in this way, we infer the population-level quantities (i.e., population-, uncertainty- and calibration-model parameters) directly, bypassing the need to perform any fits at the individual galaxy level.}
\label{fig:schematic}
\end{figure*}
Inferring the population-level parameters ${\hyper, \nuisance}$ from the hierarchical model defined in \S \ref{sec:model} is difficult for a number of reasons. Firstly, flexible (neural network) parameterizations of the population and photometric-uncertainty models mean that the number of hyper-parameters of interest is large\footnote{In this case the weights and biases of our diffusion model constitute $37,264$ free parameters characterizing the population-model.}. Secondly, there is a vast number of individual-galaxy level parameters $\{\sps, z\}_{1:N}$ that would need to be inferred and then marginalized over in a typical Bayesian analysis (using e.g.\ Markov chain Monte Carlo methods). This provides a technical challenge due to the complexity and diversity of individual galaxy SPS-parameter likelihoods (degeneracies and multimodality are commonplace), and a computational bottleneck due to the large number of SPS model calls required. Thirdly, the selection cuts introduce a high-dimensional integral into the likelihood, making it effectively intractable (see \citealp{alsing2023} for details).

Even though the likelihood is intractable, the model described in \S \ref{sec:model} defines a straightforward recipe for simulating mock catalogs, given some assumptions about the population-level parameters. This means that we may instead compare simulated catalogs to the data in a simulation-based inference setting, for example, by minimizing a suitable distance metric between model generated and real data. 

Minimizing the divergence between the predictive (model) and true data distributions is well-motivated: maximum-likelihood estimation is asymptotically equivalent to minimizing the Kulback--Leibler (KL; \citealp{kullbackleibler1951}) divergence between model and data distributions. However, the KL divergence requires evaluating the predictive distribution for the data from our model, in this case the predicted distribution of galaxy photometry for galaxies in the survey (given hyper-parameters $\hyper,\nuisance$). As discussed above, this distribution is not tractable so we seek an alternative distance metric with properties suitable for robust and efficient parameter estimation.

We estimate the population-level parameters $\hyper,\eta$ by minimizing the optimal transport (OT) distance between model generated data (catalog) $\mathbf{D} = \data_{1:N}$, and the COSMOS data $\hat{\mathbf{D}} = \hat{\data}_{1:N}$. The OT distance (also known as the Wasserstein distance or Kantorovich--Rubinstein metric, after \citealp{kantorovichrubinstein1958, vasserstein1969}) measures the divergence between two distributions from which we have samples, by computing the minimum distance required to transport one set of points onto the other, given some local metric to define distances in data space\footnote{Typically just the Euclidean or Manhattan distance.} (for a review on OT and its implementation, see \citealp{peyre2019}). Optimal transport has been widely used for parameter estimation in settings where the KL divergence is intractable \citep{peyre2019}, providing efficient and consistent estimators, which are asymptotically equivalent to maximum-likelihood estimation in the large-dataset limit in most situations\footnote{In various settings, optimal transport distance optimization is exactly equivalent to maximum-likelihood estimation \citep{rigollet2018entropic, kwon2022score}, importantly, including the generic case of fitting score-based diffusion models to data via score-matching \cite{kwon2022score}.}.

While exact calculation of the optimal transport distance is computationally complex ($\mathcal{O}{N^3\ln\,N}$; \citealp{pele2009}) and difficult to scale, the Sinkhorn divergence \citep{cuturi2013} provides a fast ($\mathcal{O}{N^2\ln\,N}$; \citealp{altschuler17, dvurechensky18}) and accurate approximation. We use the Sinkhorn divergence implemented in \texttt{pytorch} \citep{paszke2019} in the \texttt{geomloss} library \citep{feydy2019} built on \texttt{keops} \citep{charlier2021}, assuming a local Euclidean metric (2-norm) to define distances between data points.

The forward model described in \S \ref{sec:model} is stochastic: galaxy parameters are drawn from the population model distribution, calibrated model photometry is calculated and then uncertainties and noise are drawn and added, followed by application of selection cuts. In order to be able to use gradient-based optimization to minimize the OT distance between simulated and real data, we need to be able to take gradients through our simulator. To achieve this, we use a variant of the reparameterization trick \citep{kingma2013auto}, where we re-write our forward model as a sequence of deterministic steps applied to some fixed draws from a base density (which are kept fixed for the purpose of estimating gradients). In this sense, our forward model can be written as the following sequence of steps:
\begin{enumerate}
    \item Draw base random variates for the population-model, uncertainty-model, and noise-model:
    \begin{align}
        &\bm{u}_{1:M}\sim \mathcal{N}(0, 1), \;[\text{population-model base draws}] \nonumber \\
        &\bm{s}_{1:M}\sim \mathcal{N}(0, 1), \;[\text{uncertainty-model base draws}] \nonumber \\
        &\bm{n}_{1:M}\sim \mathcal{T}_2, \;[\text{noise-model base draws}] \nonumber
    \end{align}
    where $\mathcal{T}_2$ is the standard-$t$ distribution with two degrees-of-freedom (Equation \ref{eq:data}), and $M$ is the number of mock galaxies to generate (which should be larger than the target (selected) catalog size $N$);
    \item Pass base samples $\bm{u}_{1:M}$ to the population-model (score-based diffusion model) to generate draws of galaxy parameters $\sps_{1:M}$, by solving the ODE in Equation \eqref{eq:ode} (given the current values of the population-model parameters $\hyper$);
    \item Compute calibrated photometry $\flux_{1:M}$ for each galaxy assuming the SPS and calibration models (Equation \ref{eq:calibration}, given the current values of the data-model parameters $\nuisance$);
    \item Pass base samples $\bm{s}_{1:M}$ and the model fluxes $\flux_{1:M}$ through the uncertainty model (Equation \ref{eq:draw_unc}) to generate photometric noise variances for each galaxy $\noise_{\mathrm{P},1:M}$ (given the current values of the uncertainty-model parameters $\unc$);
    \item Compute additional uncertainty contributions $\noise_{\mathrm{EM},1:M}$ due to emission-lines (Equation \ref{eq:emline-unc}), and add in quadrature to get total uncertainties $\noise_{1:M}^2 = \noise_{\mathrm{P},1:M}^2 + \noise_{\mathrm{EM},1:M}^2$;
    \item Pass base samples $\bm{n}_{1:M}$ and the model photometry to the noise model (Equation \ref{eq:data}) to generate noisy mock photometry, $\data_{1:M} = \flux_{1:M} + \noise_{1:M}\odot\bm{n}_{1:M}$;
    \item Apply selection cuts (and trim the number of retained objects to $N$ if necessary) to give a mock catalog $\Data(\bm{u}, \bm{s}, \bm{n}|\hyper, \nuisance) = \{\data_{1:N}, S_{1:N}=1\}$ of the desired size, $N$.
\end{enumerate}
The objective function for minimization is then given by:
\begin{align}
    \mathcal{L}(\hyper, \nuisance) = \mathcal{W}_2[\Data(\bm{u}, \bm{s}, \bm{n}|\hyper, \nuisance), \hat{\Data}],
\end{align}
where $\mathcal{W}_2$ denotes the OT distance (assuming a local Euclidean metric), $\Data(\bm{u}, \bm{s}, \bm{n}|\hyper, \nuisance)$ is the simulated catalog and $\hat{\Data}$ the COSMOS catalog. By keeping the base random drawn from step 1 fixed, the simulated catalog (and hence OT distance) are deterministic functions of the parameters $\hyper, \nuisance$, so that we can take gradients and perform gradient-based optimization. This scheme is summarized in Figure \ref{fig:schematic}.
\subsection{Initialization and training}
The calibration model parameters (characterizing the zero-points and emission-line corrections) are initialized following the Bayesian hierarchical calibration approach presented in \cite{leistedt2023hierarchical}: cross-matching with data from \texttt{zCosmos-bright} \citep{Lilly_2007}, \texttt{DEIMOS} \citep{Hasinger_2018}], and \texttt{C3R2} \citep{Masters_2017, Masters_2019, Stanford_2021} yields 12,473 objects with spectroscopic redshifts available, in the range $0 < z < 2$. This lifts degeneracies between SPS parameters and makes the calibration model parameters very well constrained by the data. Simultaneous optimization of all parameters converges easily, with the SPS parameter uncertainties having negligible effect on the result (see \citealp{leistedt2023hierarchical} for more details).

To initialize the population model, we perform an initial maximum aposteriori (MAP) estimation of the SPS parameters for each galaxy in the COSMOS sample, and pre-train the diffusion model on that ensemble of MAP estimates via denoising score-matching. This provides a reasonable initialization for the population model to avoid a long burn-in phase based on the more expensive optimal transport objective.

The uncertainty model network is initialized as follows. The initial MAP estimates for the SPS parameters (and initialized calibration-model parameters) provide estimates of the true (denoised) photometry for each galaxy in the COSMOS sample. This provides a catalog of uncertainties and associated (denoised) photometry $\{\bm{\sigma}_\mathrm{P}, \flux\}$, on which we can train our conditional estimator for $P(\bm{\sigma}_\mathrm{P} | \flux; \unc)$ by minimizing the negative log-likelihood loss:
\begin{align}
    \mathcal{L}(\unc) = -\sum_{i=1}^{N_\mathrm{train}} \ln P(\bm{\sigma}_{\mathrm{P,}i} | \flux_i; \unc).
\end{align}
This provides a reasonable initialization for the uncertainty model, after which $\unc$ is kept free in the final fitting procedure.

OT optimization is then performed with \texttt{Adam} \citep{kingma2014adam} with a learning rate of $10^{-4}$, until the distance ceases to improve for twenty epochs. All of the population-level hyperparameters are kept free in the fitting process, including the zero-points and emission line corrections.

We compute the OT objective between both the synthetic and real magnitudes, and separately between the synthetic and real colors\footnote{$25$ adjacent-band colors.}, and sum them. We find that this improves the ability of the fitted model to reproduce both the colors and magnitudes faithfully.
\subsection{Discussion}
The model fitting scheme described above has a number of advantages over existing methods.

Firstly, we target the hyper-parameters (describing population- and data-model) directly, completely bypassing the need to infer the properties of each individual galaxy in the sample (in contrast to eg., MCMC-based approaches). This provides a significant advantage in computational cost and scalability when population-level inference is the main goal.

Secondly, by jointly inferring the population- and data-model parameters together in a self-consistent fashion, we are able to use the data to ``self-calibrate'' any unknown nuisance parameters (e.g., calibration and noise-model parameters, etc). This will result in more robust inferences compared to the traditional approach of estimating and fixing nuisance parameter values prior to the main analysis.

While our fit to COSMOS data necessarily includes the $r < 25$ selection cut, our inference pipeline explicitly includes (and corrects for) that selection: the target population model is therefore a description of the underlying galaxy population that is not subject to selection effects. The resulting population model can therefore be straightforwardly utilized for prediction (and like-for-like comparison) for different surveys, simply by applying the noise characteristics and selection appropriate for that survey in a forward modeling context. This point is critical for application to cosmological inference from broad-band galaxy surveys, where we require a well-calibrated population model that is able to make faithful predictions for those deep, broad-band data. Note that while our method properly corrects for selection, it is not designed to extrapolate more than a few noise standard deviations below the flux-limit (where there is no data to constrain the population model). Therefore, application of the calibrated population model should be limited to surveys with similar or shallower depths.

Our forward modeling-based approach is also well suited to principled validation on the basis of predictive performance. This is in contrast to typical galaxy evolution and cosmology analyses, where population-level inferences are drawn, but little assessment of prediction quality (in data-space) is done. In a companion paper \citep{thorp24}, we present a model validation approach for the simulation-based inference setting, based on quantile--quantile (QQ) and probability--probability (PP) plots (\citealp{wilk68}; see \citealp{eadie23} for an astronomy example). A further complication is the question of comparing models. Again, the most popular approaches in astrophysics and cosmology are typically applied at the level of the parameter posterior (i.e.\ via the Bayesian evidence; although use of posterior predictive scores is growing, see e.g.\ \citealp{feeney19, abbott19, rogers21, setzer23, mcgill23, welbanks23, nixon23}). In our simulation-based approach, we can readily interrogate competing models based on their ability to reproduce observed data.

The simulation-based inference scheme described above currently provides a point estimate for the population-level parameters. Statistical uncertainties on the estimated parameters could be obtained by bootstrapping, or training ensembles of models with different initializations (e.g., \citealp{li2024popsed}). However, we expect uncertainties to be dominated by systematic rather than statistical errors (due to e.g., photometric calibration; see \S~\ref{sec:data-space}).
\section{Results}
\label{sec:results}
\begin{figure*}
\centering
\includegraphics[width = \linewidth]{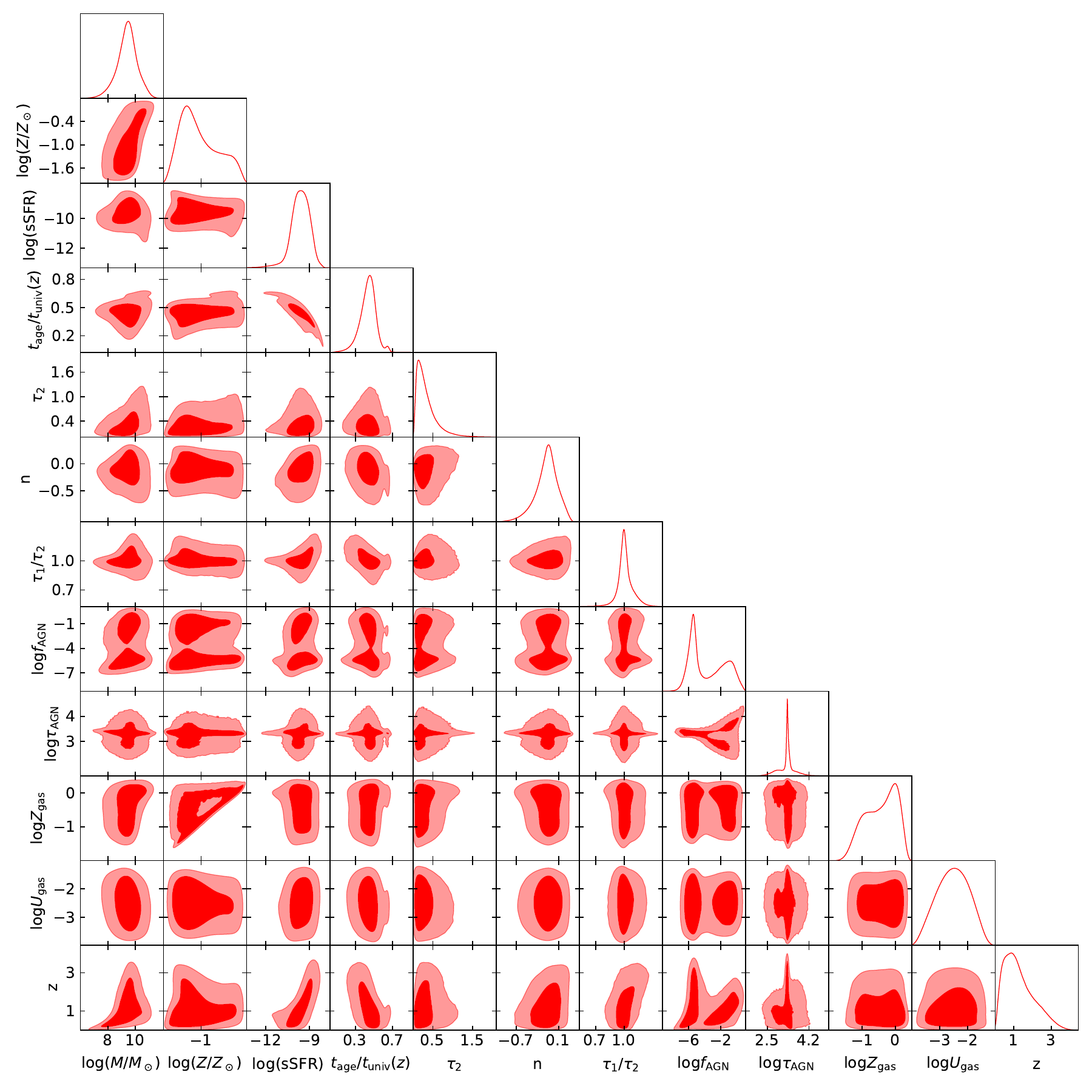}
\caption{1- and 2-d marginals of the SPS parameters, predicted by our galaxy population model. SPS parameters shown comprise (see also Table \ref{tab:notation}): stellar mass $(M/M_\odot)$ and metallicity $(Z/Z_\odot)$; specific star-formation rate $\mathrm{sSFR}$; mass-weighted age $t_\mathrm{age}/t_\mathrm{univ}(z)$, optical depths of the diffuse ($\tau_2$) and birth-cloud ($\tau_1$) dust screens; the index of the dust-attenuation law $n$ (relative to \citealp{calzetti2000}); the fraction of the bolometric luminosity from AGN $f_\mathrm{AGN}$; the optical depth of the AGN torus $\tau_\mathrm{AGN}$; the gas metallicity and ionization parameter $Z_\mathrm{gas}$ and $U_\mathrm{gas}$; and redshift $z$. The star-formation rate and age are derived quantities; we assume a non-parametric (piecewise-constant) model for the SFH (see \S~\ref{sec:sps-model}).}
\label{fig:triangle}
\end{figure*}
In this section we present the key results from our fitted forward model. Our model predictions in data-space are validated against the COSMOS sample in \S \ref{sec:data-space}, and the fitted values of the calibration-model parameters (zero-points and emission-line corrections) are given in Tables \ref{tab:zps} and \ref{tab:emlines}. While most of the emission-line corrections are at the percent level or less, ten of the included bands get $\gtrsim 10\%$ or more (and up to $50\%$ in some cases). We report that emission-line calibration was essential to obtain physically reasonable population-model constraints on the fundamental-metallicity relation (gas-metallicity vs. SFR; Figure \ref{fig:FMR}), and AGN (Figure \ref{fig:triangle}).

The 1- and 2-d marginals for the fitted population-model are summarized in Figure \ref{fig:triangle}. Since we assumed a flexible parameterization for the population model, it is designed to capture the complete web of complex dependencies between galaxy characteristics and how those evolve over cosmic time. While some of this structure is already visible in Figure \ref{fig:triangle}, we present our model predictions in light of commonly studied relationships and quantities in \S \ref{sec:redshift}-\ref{sec:gas}: the redshift distribution \S \ref{sec:redshift}; mass-function \S \ref{sec:mass-function}; mass-metallicity and fundamental relations \S \ref{sec:mass-met}-\ref{sec:FMR}; dust versus mass and SFR \S \ref{sec:dust}; and gas ionization versus SFR \S \ref{sec:gas}. Constraints on AGN are briefly discussed in \S \ref{sec:AGN}. Note that while our model corrects for selection, our flexible population-model parameterization is not designed to extrapolate far below the flux-limit of the sample (where the data has no constraining power): this leads to the apparent turnover at low masses (high redshifts) in Figure \ref{fig:triangle}. 

Direct quantitative comparison with previous work for the relations presented in \S \ref{sec:redshift}-\ref{sec:gas} is not always straightforward. This work represents the first time it has been possible to jointly infer the full set of galaxy parameter dependencies, while accounting for SPS-parameter degeneracies, self-calibrating the data- and calibration-model, and correcting for selection. It is therefore non-trivial to present like-for-like comparisons with previous studies with different SED or data-modeling assumptions, different assumed priors or specific parametric forms for scaling relations, and differing selection effects. For these reasons, in \S \ref{sec:redshift}-\ref{sec:gas} we focus primarily on presenting a broad physical interpretation of our results, with qualitative comparison to the (most-comparable) literature where appropriate, and to template-based parameter estimates in some cases as a sanity check. We leave detailed comparison with the literature and implications for galaxy evolution to future work.

\subsection{Data-space comparisons}
\label{sec:data-space}
\begin{figure*}
\centering
\includegraphics[width = \linewidth]{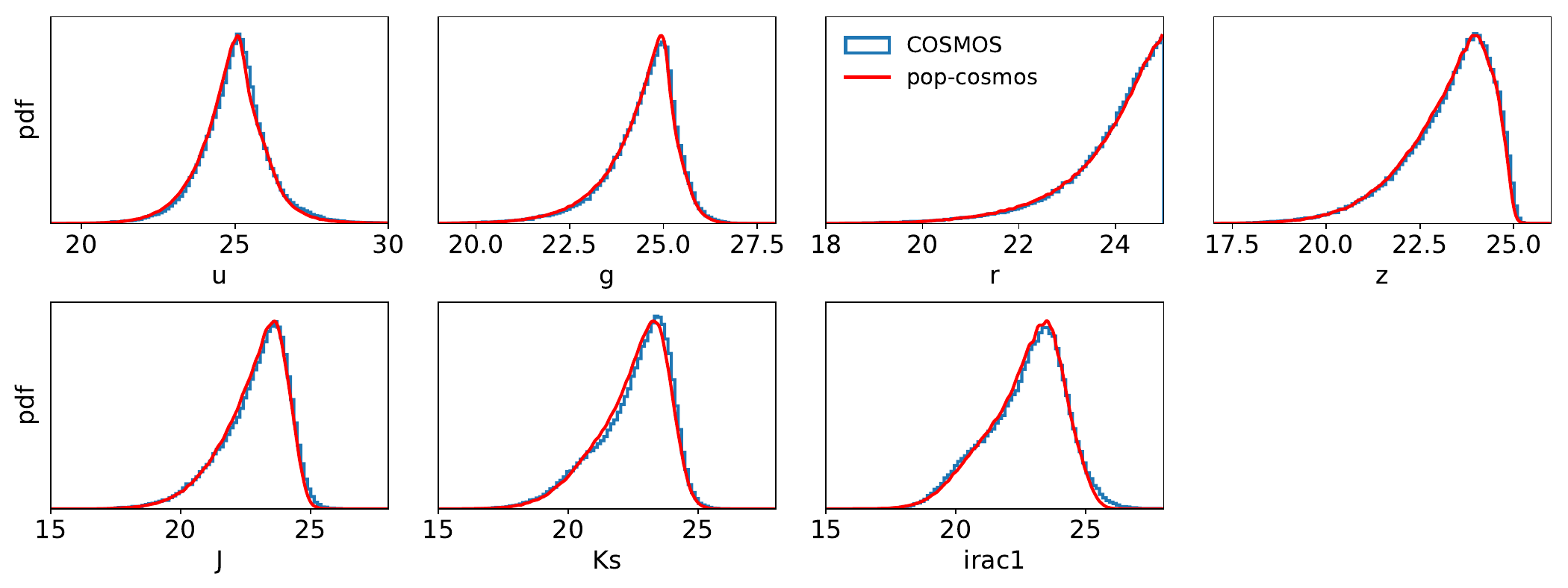}
\caption{Comparison of the marginal distributions for the magnitudes predicted by our model (red), versus the COSMOS data (blue). Comparison is shown in the observed data-space, i.e., for $r<25$ selected galaxies and with photometric noise and calibration included (to ensure like-for-like comparison with the data). We show a subset of the $26$ bands, spanning the full wavelength range.}
\label{fig:magnitude_marginals}
\end{figure*}

\begin{figure*}
\centering
\includegraphics[width = \linewidth]{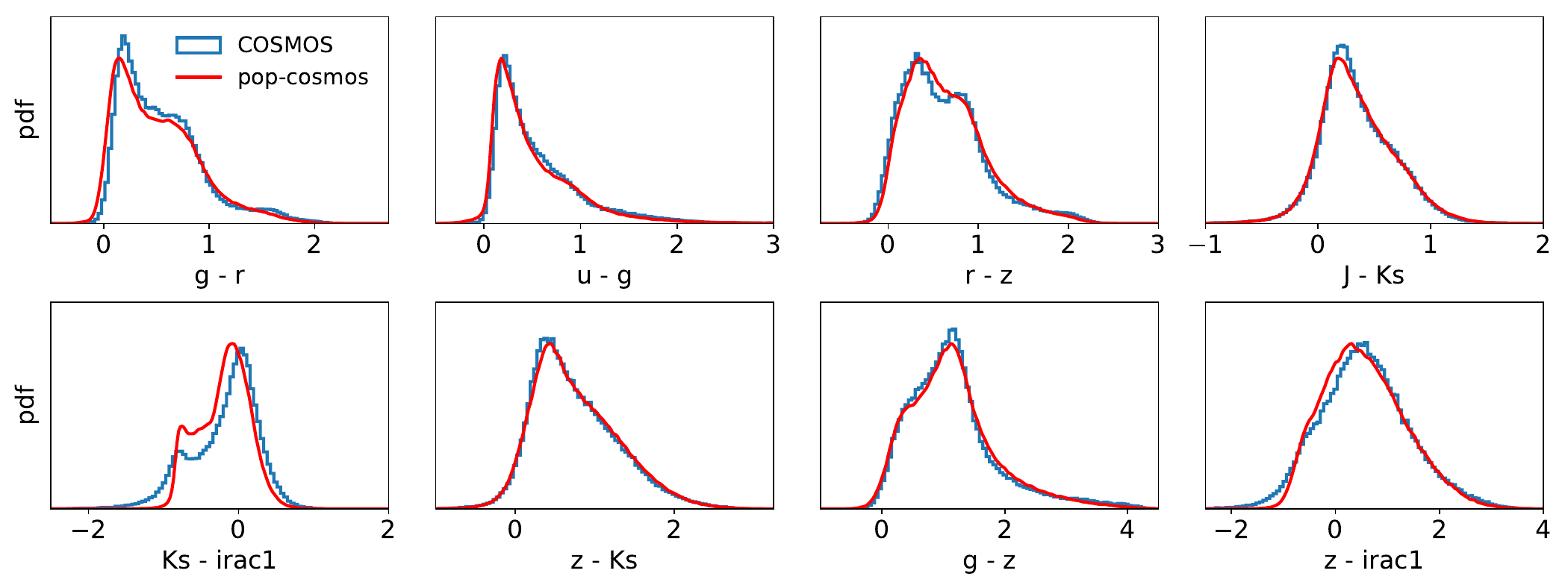}
\caption{Comparison of the marginal color distributions predicted by our model (red), versus the COSMOS data (blue). Comparison is shown in the observed data-space, i.e., for $r<25$ selected galaxies and with photometric noise and calibration included (to ensure like-for-like comparison with the data). We show a subset of key colors following \cite{weaver2020}.}
\label{fig:color_marginals}
\end{figure*}

\begin{figure*}
\centering
\includegraphics[width = \linewidth]{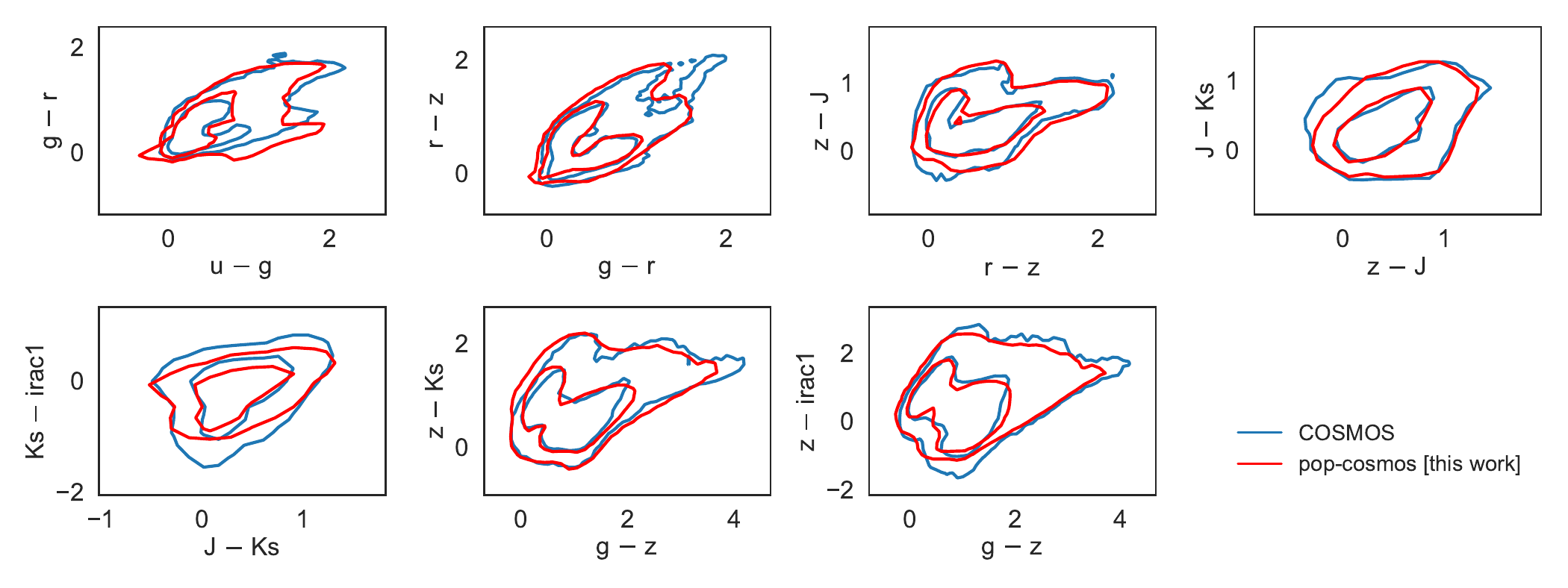}
\caption{Color-color diagrams comparing the COSMOS data (blue) to predictions from our trained forward model (red), in the observed data-space (i.e., with photometric noise and calibration included in the model predictions to ensure like-for-like comparison with the data). Contours show 68 and 95 per cent levels. We show a subset of key colors following \cite{weaver2020}.}
\label{fig:color_color}
\end{figure*}

Comparisons of our fitted model predictions to the COSMOS data in magnitude- and color-space are shown in Figures \ref{fig:magnitude_marginals}--\ref{fig:color_color}. To ensure a like-for-like comparison, Figures \ref{fig:magnitude_marginals}--\ref{fig:color_color} compare our model predicted distributions for noisy, calibrated ($r < 25$ selected) photometry against the equivalent COSMOS data. We focus on a subset of key bands and colors, spanning the full wavelength range and key color-space features, following \cite{weaver2020}.

Our model achieves excellent agreement in the magnitude marginals (Figure \ref{fig:magnitude_marginals}); this is not unexpected, since the magnitude marginals are mostly dominated by the shape of the mass function and volume effects, which should be easily-captured by the model.

The predictive distribution of galaxy colors on the other hand is a rich probe of galaxy evolution physics. The ability of our model to faithfully reproduce the color-color distribution underpins our confidence in the model predictions, and accurate characterization of galaxy colors as a function of redshift is crucial for predicting redshift distributions for cosmological analyses (e.g.\ \citealp{alsing2023}). In Figures \ref{fig:color_marginals} and \ref{fig:color_color} we see that our model reliably reproduces the color-color distributions of COSMOS galaxies, including fine structure (e.g.\ related to star-forming and quiescent concentrations). 

The largest discrepancies (at the level of $0.05-0.1$ magnitude color offsets) are seen in ($K_s-\texttt{irac1}$) and ($g-r$). These small biases are likely explained by residual (un-modeled) systematics in the COSMOS data. \cite{weaver2020} performed a detailed comparison of the \texttt{Farmer} and \texttt{Classic} versions of the COSMOS catalogs, with different approaches to the photometric extraction. They reported the largest unexplained systematic differences between the two catalogs' photometry in the \texttt{irac1}, $g$ and $u$ bands (figure 8 of \citealp{weaver2020}), with ($K_s- \texttt{irac1}$), ($g-r$) and ($z-J$) being the most affected colors (figure 9 of \citealp{weaver2020}). Discrepancies between our model predictions and the \texttt{Farmer} data are less than the systematic differences between \texttt{Farmer} and \texttt{Classic} in all bands and colors. It is therefore likely that any modest differences between model and data seen in Figures \ref{fig:color_marginals} and \ref{fig:color_color} are dominated by residual systematics in the COSMOS photometry. This makes a strong case for pursuing further improvements to the photometric data-modeling and extraction for COSMOS data in future\footnote{Conversely, the fact that our flexible population- and SPS-models are able to avoid simply ``overfitting" to residual un-modeled systematics in the photometry is encouraging. This is because the model is physics-guided, and helps build confidence in the model predictions.}.

In a companion paper \citep{thorp24}, we will present further validation of our calibrated model in data-space, based on quantile--quantile (QQ) and probability--probability (PP) plotting.

\subsection{Redshift distribution}
\label{sec:redshift}
Accurate prediction of the redshift distributions for ensembles of (photometrically-selected) galaxies is of critical importance in constraining cosmology from weak lensing surveys (see e.g., \citealp{newman2022} for a review). Redshift distributions are a key ingredient in predicting lensing and clustering statistics from data, and have significant degeneracies with key cosmological parameters of interest. This leads to very stringent requirements on their accuracy; for imminent Stage IV surveys such as LSST \citep{lsst}, biases on inferred redshift distributions must not exceed $0.001(1+z)$ (in the mean redshift; \citealp{mandelbaum2018}).

Forward modeling has emerged as a promising avenue for obtaining accurate redshift distributions for deep broad-band imaging surveys, where sufficient spectroscopic calibration data are not available \citep{alsing2023}. These approaches rely on accurate modeling of the galaxy population, with calibration to deep flux-limited samples such as COSMOS (as in this work) expected to provide key baseline constraints.

\begin{figure}
\centering
\includegraphics[width = 8.5cm]{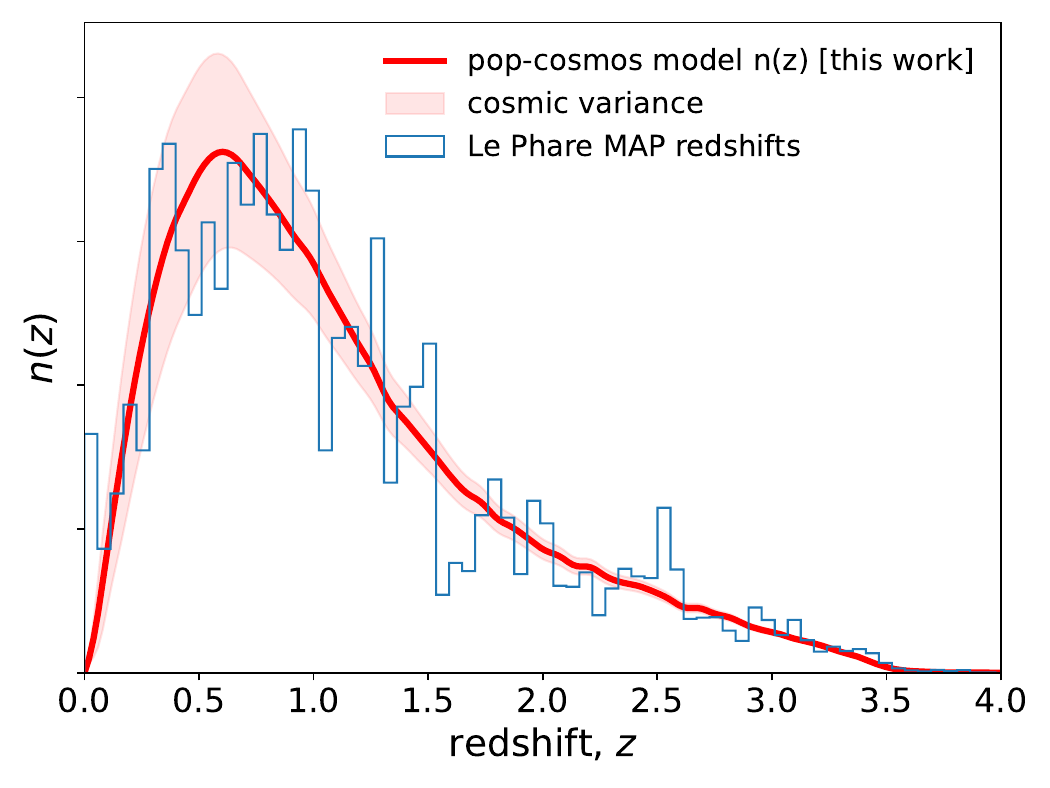}
\caption{Photometric redshift distribution predicted by our forward model (red curve), with estimated uncertainties due to cosmic variance (red envelope). The blue histogram shows redshift estimates for COSMOS using \texttt{LePhare} \citep[from][]{weaver2020}.}
\label{fig:nz}
\end{figure}

In Figure \ref{fig:nz} we show our predicted galaxy redshift distribution $n(z)$ (given the photometric cuts described in \S \ref{sec:data}), and compare to photometrically-derived redshift estimates from \texttt{LePhare} \citep{weaver2020}. Cosmic variance is estimated following the recipe in \citet{moster2011}\footnote{The cosmic variance estimation is performed using redshift bins of $\Delta z = 0.05$}.

The predicted $n(z)$ is broadly in good agreement with the \texttt{LePhare} redshift estimates, with two notable discrepancies. Firstly, the \texttt{LePhare} redshifts exhibit an unphysical build-up of low or zero redshift galaxies. This is a commonly observed feature in template-based photo-$z$ estimation, where some fits get driven to the prior boundary at $z=0$, while the assumed redshift prior does not go to zero at the boundary to penalize them appropriately \citep{hildebrandt2012cfhtlens}\footnote{The common practice of using redshift priors that do not go to zero at $z=0$ was introduced in \cite{hildebrandt2012cfhtlens} as an ad hoc modification that was observed to reduce the bias in template-based redshift estimates at low redshift. However, it comes at a cost of (un-physically) allowing some template fits to be driven up against the prior boundary at $z=0$.}. Second, the \texttt{LePhare} redshift histogram exhibits clustering above $z>1$ over-and-above the expected clustering due to cosmic variance. This behavior is commonplace in template-based photo-$z$ methods, where redshift point estimates have a tendency to cluster around specific values (owing to the limited fidelity with which finite or interpolated template-sets can describe real galaxy SEDs). In contrast, our predicted $n(z)$ has the physically correct behavior of going to zero at $z=0$, and does not exhibit spurious structure above $z>1$. Conversely, since our generative model does not include galaxy clustering (present in the COSMOS sample at $z\lesssim 1$), and is only calibrated to galaxy colors (which are expected to be very weakly sensitive to clustering), our model implicitly learns the underlying (mean) $n(z)$, as desired.

We expect the calibrated \texttt{pop-cosmos} model to provide an improved population-model for predicting redshift distributions for cosmological surveys \citep{alsing2023}. We will present \texttt{pop-cosmos} enabled redshift distribution estimation for KiDS data in a companion paper (Loureiro et. al., in prep.).

While we do not present individual galaxy redshift estimates here, our calibrated population model can also provide an improved prior for SPS-based photo-$z$ estimation. We are investigating the utility of \texttt{pop-cosmos} for redshift estimation for individual galaxies in a companion paper \citep{thorp24_mcmc}.
\subsection{Galaxy stellar mass-function}
\label{sec:mass-function}
Galaxies build up stellar mass through a combination of in-situ star formation and mergers. Modeling how galaxies grow is a major ongoing challenge, involving processes that span a wide range of scales (from stellar to cosmological; see e.g., \citealp{somerville2015physical} for a review). Observations of the stellar mass function and its redshift evolution hence provide an important constraint on models of galaxy formation and evolution \citep{marchesini2009, ilbert2013, Muzzin2013a, moustakas2013, tomczak2014galaxy, grazian2015galaxy, song2016evolution, davidzon2017cosmos2015, wright2018gama, leja2020MF, weaver2023MF}. In the context of photometric redshift estimation, accurate characterization of the mass function is also essential for obtaining accurate redshifts.

\begin{figure*}
\centering
\includegraphics[width = 17.5cm]{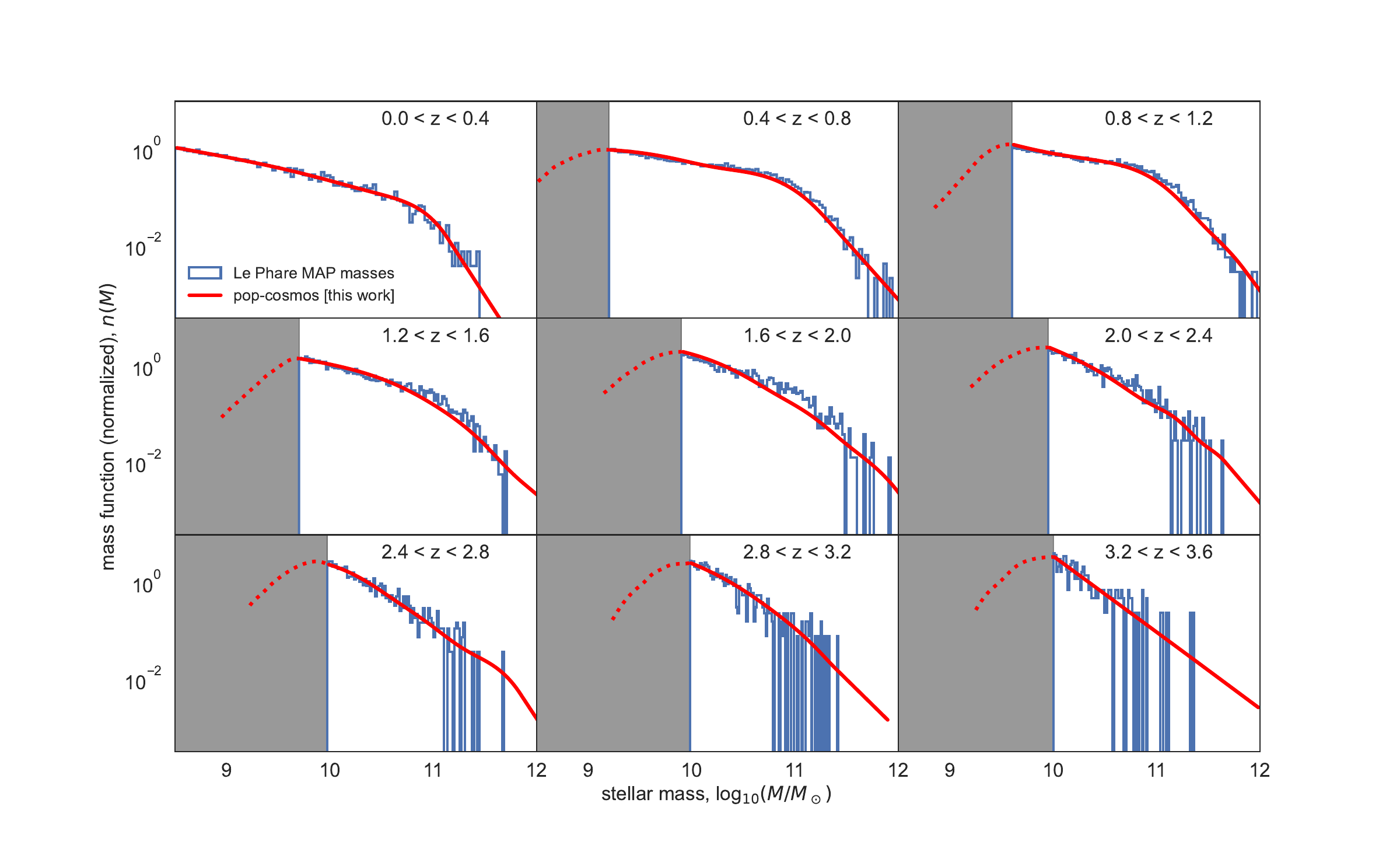}
\caption{Predicted galaxy stellar mass functions in redshift bins of widths $\Delta z=0.4$. Our forward model predictions are shown in red (dotted in the incomplete region), with the \texttt{LePhare} mass distributions shown as blue histograms. To ensure a like-for-like comparison with \texttt{LePhare}, we show the mass function for galaxies with an $r$-band magnitude cut $r<25$. Gray shaded regions indicate the region where the selected sample plotted begins to become incomplete (see footnote \ref{foot:completeness}).}
\label{fig:mass_function}
\end{figure*}

The stellar mass function derived from our model is shown in Figure~\ref{fig:mass_function}\footnote{\label{foot:completeness}The completeness limits shown in Figures \ref{fig:mass_function}- \ref{fig:FMR} are estimated by visual inspection of the turnover of the mass function (for $r<25$ selected galaxies); they are intended as a visual guide only. Completeness limits do not explicitly appear anywhere in our analysis, and we hence did not make a detailed quantitative evaluation of them.}. The closest study for comparison is \citet{weaver2023MF}, who estimate the stellar mass function from COSMOS2020 data based on the \texttt{LePhare} mass estimates. To simplify the comparison (eliminating any differences in data and modeling assumptions) in Figure~\ref{fig:mass_function} we compare directly to the \texttt{LePhare} masses on which the \citet{weaver2023MF} measurement is based.

We achieve good agreement with the \texttt{LePhare} masses over the entire redshift range, and predict a number of key features in the mass function. We find a steepening of the low-mass slope with redshift, a buildup of galaxies around $10^{11}M_\odot$ below $z < 1.2$ (leading to the observed ``bump" in the mass function at low and intermediate redshifts), and little or no redshift dependence of the location of the knee of the mass-function. We also note relatively little evolution in the shape of the mass function at $z\lesssim 1.5$. These features are in good agreement with previous observations (including previous COSMOS analyses: \citealp{ilbert2013,davidzon2017cosmos2015,weaver2023MF}).

Comparison to other recent measurements such as \citet{leja2020MF} are non-trivial due to differing modeling assumptions; we leave broader comparisons to future work.
\subsection{Star-forming sequence}
\label{sec:SFS}
The star-forming sequence (SFS) characterizes the relationship between star formation rate (SFR) and stellar mass, with galaxies generally forming most of their mass either on \citep{leitner2012}, or passing through \citep{abramson2015}, the star-forming sequence. Measurements of the SFS hence provide an important probe of galaxy evolution and cosmic star-formation history \citep{daddi2007, noeske2007, karim2011, rodighiero2011, whitaker2012, whitaker2014, speagle2014, renzini2015, schreiber2015, tomczak2016, leslie2020, leja2021SFS}.

\begin{figure*}
\centering
\includegraphics[width = 17.5cm]{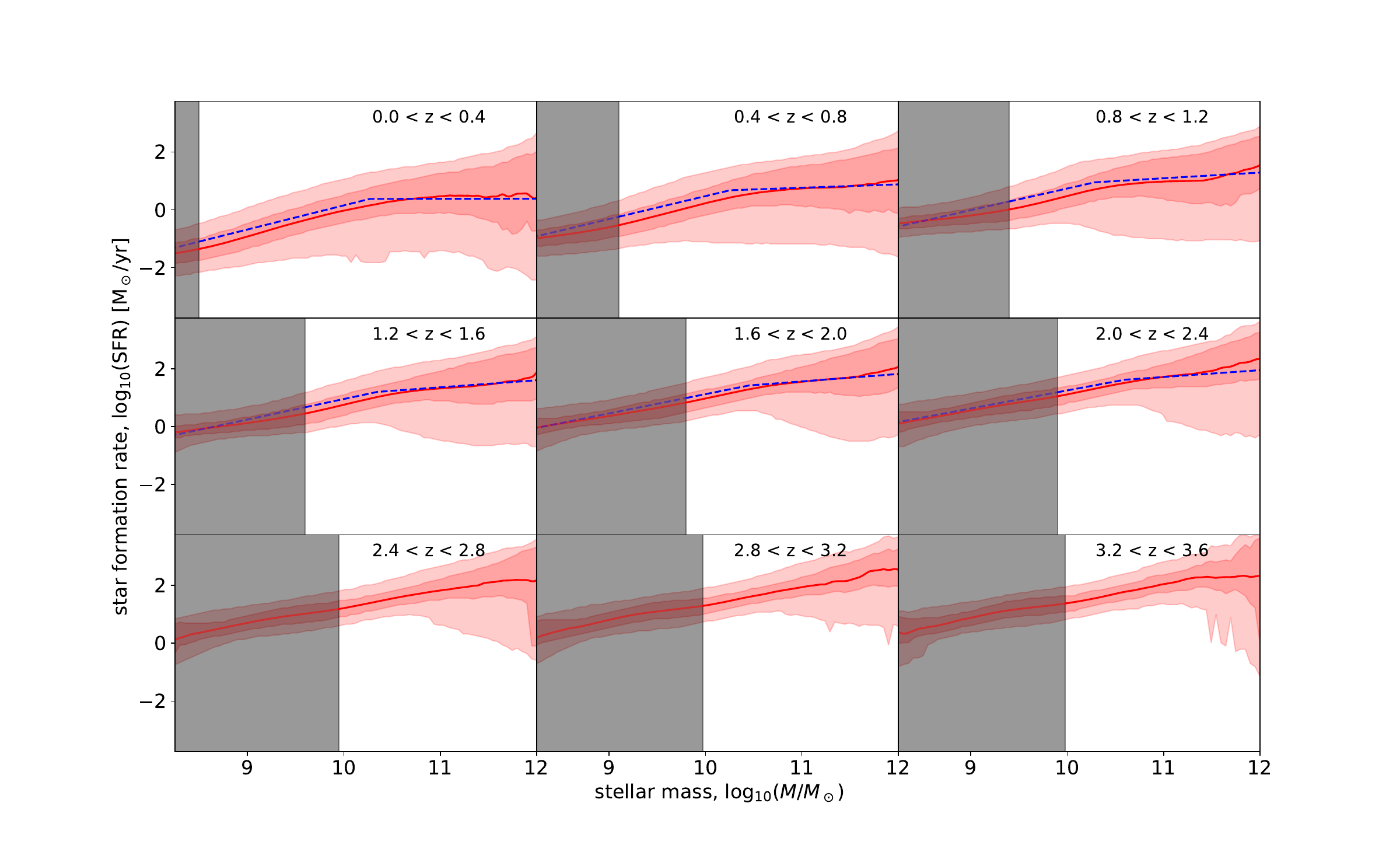}
\caption{The star-forming sequence predicted by our forward model. The red line shows the median of our predictions of $\log10(\mathrm{SFR})$, with the red shaded regions showing the 68 and 95\% intervals of the conditional distribution at a given mass (estimated in a rolling mass window). The gray shaded region shows the estimated mass completeness limit in each redshift bin (see footnote \ref{foot:completeness}). All subsequent figures follow the same plotting scheme unless noted. The blue dashed lines show the inferred SFS from \citet{leja2021SFS}. Note that the \citet{leja2021SFS} predictions are for the logarithm of the mean SFR.}
\label{fig:SFS}
\end{figure*}

Figure \ref{fig:SFS} shows the inferred relationship between SFR, stellar mass, and redshift, from our population model. We compare to the measured SFS from \cite{leja2021SFS}, which is based on COSMOS-2015 and 3D-HST photometry. This comparison is chosen because it is the most similar to our analysis; they model the SFS for star-forming and quiescent galaxies together (as in our work, rather than selecting star-forming galaxies only), the datasets used have some commonality, and the SPS modeling assumptions are similar.

Our recovered SFS is in good agreement with the measurement from \cite{leja2021SFS}. We find a similar slope of the SFS at both low and high masses, flattening of the SFS at higher masses, steepening of the high-mass slope as a function of redshift, and a negative skewness at the high-mass end owing to the increasing presence of massive quiescent galaxies at higher masses. The small offset in normalization at low masses is mostly due to the broken power-law from \cite{leja2021SFS} modeling the log of the mean SFR, whereas for our model we show the median log SFR. We would also expect some modest quantitative differences due to the differing galaxy samples used, treatment of selection effects, and modeling assumptions. We note that our inferred SFS extrapolates sensibly into the regime where the COSMOS data are incomplete or lacking (Figure \ref{fig:SFS}, grey bands).

The majority of observational studies have focused on characterizing the SFS for star-forming galaxies only, since those are the galaxies which are actively forming mass. However, more recently it has been shown that the method of identifying star-forming galaxies leads to systematic differences in the inferred SFS (of up to $0.5$ and $0.2$ dex in normalization and width respectively; \citealp{leja2021SFS}), owing largely to the fact that the galaxy population cannot be cleanly split into ``star-forming" and ``quiescent" samples based on SFR (ie., the distribution of SFR is not strongly bimodal at most masses and redshifts: see e.g., \citealp{leja2021SFS}). We emphasize that, in the spirit of \cite{leja2021SFS}, the SFS prediction from our model presented in Figure \ref{fig:SFS} includes all galaxies in the flux-limited COSMOS sample (not only star-forming galaxies).

\subsection{Mass-metallicity-redshift}
\label{sec:mass-met}
The chemical enrichment of galaxies is driven by two main processes: successive generations of massive stars produce metals via nucleosynthesis and return them to the interstellar medium at the end of their lives; at the same time, outflows driven by starburst winds or AGN feedback result in ejection of metal-enriched gas into the intergalactic medium, while inflows can bring metal-poor gas in. The interplay of these processes results in a relationship between stellar mass, stellar- and gas-phase metallicities, and star-formation rate. The observed mass-metallicity and fundamental metallicity (mass-gas metallicity-SFR) relations hence provide key observational probes of galaxy evolution \citep{tremonti2004origin, maiolino2008amaze, mannucci2009lsd, lara2010fundamental, yates2012, lara2013galaxy, andrews2013, nakajima2014, yabe2015, salim2014, salim2015, kashino2016, cresci2019, cullen2021nirvandels, curti2020, bellstedt21, sanders2021mosdef, thorne2022devils}.

\begin{figure*}
\centering
\includegraphics[width = \linewidth]{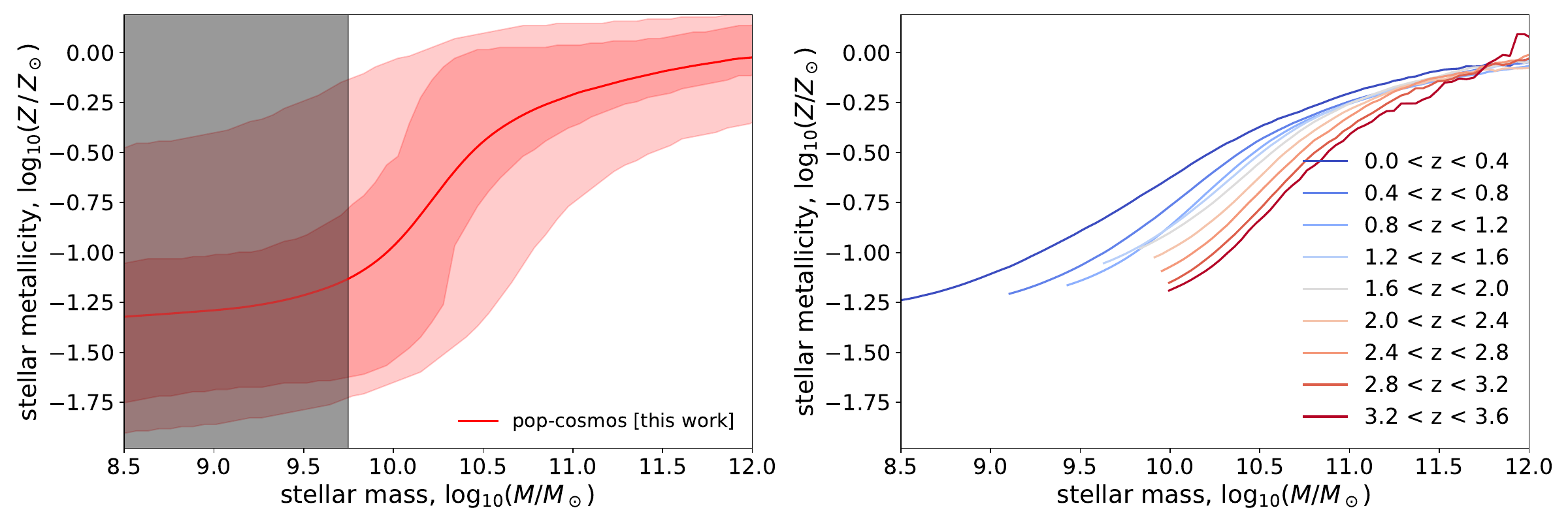}
\includegraphics[width = \linewidth]{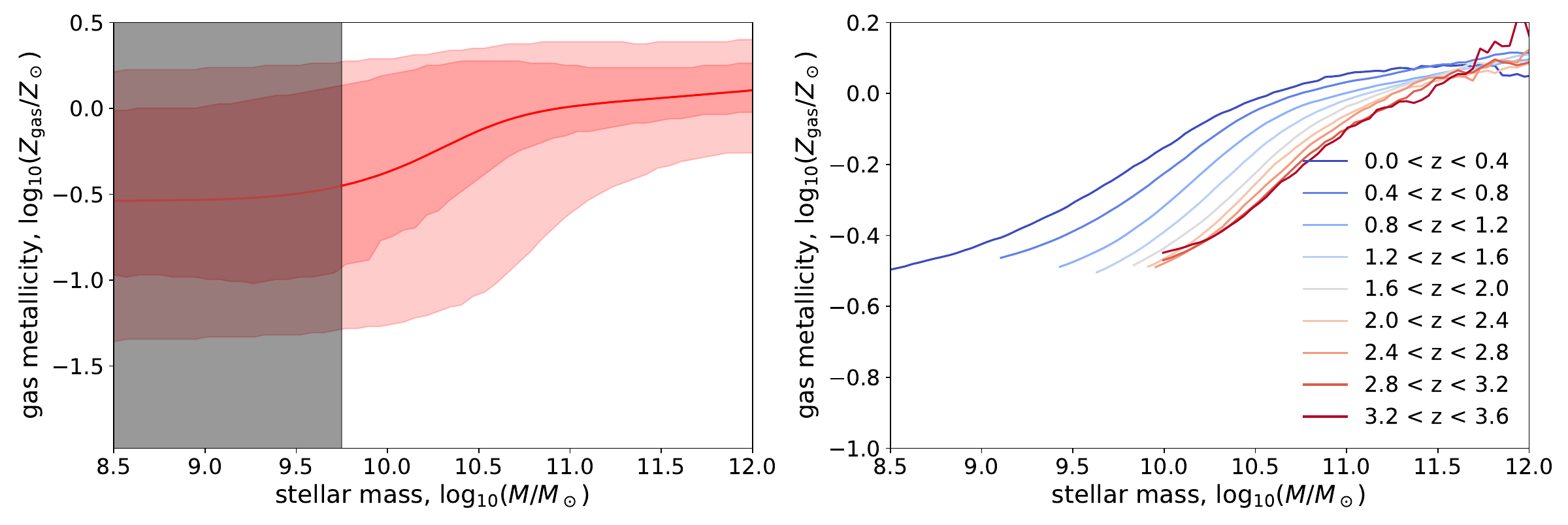}
\caption{Upper left panel: predicted stellar metallicity vs.\ mass relation. Upper right panel: median predicted mass--metallicity relation in redshift bins of width $\Delta z=0.4$. Lower left and right panels are the same, but for gas metallicity. Grey bands indicate where the COSMOS sample becomes incomplete (see footnote \ref{foot:completeness}).}
\label{fig:mass-met}
\end{figure*}

In Figure \ref{fig:mass-met} (upper panels) we show the predicted mass-stellar metallicity relation from our population model, averaged over redshift (left panel), and as a function of redshift (right panel). The shape of the mass-metallicity relation is in excellent agreement with local measurements from SDSS at low redshift (e.g., \citealp{gallazzi05})\footnote{Comparison of the normalization of the mass-metallicity relation relative to \cite{gallazzi05} is non-trivial: the metallicity measurements used in \cite{gallazzi05} are known to be biased high due to the fact that fibre spectra are used. Nonetheless, our result is consistent in normalization with \cite{gallazzi05} to within $0.3$ dex, well-within expected variations between different approaches to metallicity calibration \citep{kewley2008metallicity}.}. We find that the slope of the mass-metallicity relation at lower masses steepens by a factor of $\simeq 2$ between $z=0$ and $z=3.5$, with the trend decreasing as the mass-metallicity relation flattens off at higher masses.

In the lower panels of Figure \ref{fig:mass-met} we show our population model predictions for the mass-gas metallicity relation averaged with respect to redshift (left), and as a function of redshift (right). The shape and redshift evolution is in good qualitative agreement with recent measurements (e.g., \citealp{bellstedt21, thorne2022devils}). We find the slope of the mass-gas metallicity relation at lower masses steepens by a factor of $\simeq 2$ between $z=0$ and $z=3.5$, with the median gas metallicity at $10^{10}M_\odot$ decreasing by around $0.4$ dex over the same redshift range. This is roughly consistent with recent measurements from GAMA data \citep{bellstedt2020galaxy}, which shows around $0.6$ dex decrease in the median gas-metallicity over a similar redshift range.

The normalization of the recovered mass-gas metallicity relation (bottom row of Figure \ref{fig:mass-met}) is higher than for the mass-stellar metallicity relation (top row of Figure \ref{fig:mass-met}). This is expected, since under our assumed SPS model the gas metallicity represents the present-day metallicity of the ISM, while the stellar-metallicity parameter is a proxy for the light-weighted average metallicity among the stellar population (which includes older stars).

Note that for both stellar and gas metallicity, in the regime where the COSMOS data is lacking (grey bands in Figure \ref{fig:mass-met}) the extrapolation of our model predictions show a flattening of the mass-metallicity relations, while from observations it is expected to continue downwards (e.g., \citealp{kirby2013universal}). Our model is not designed to extrapolate very far into the regime where the data is lacking; additional constraints may be needed to improve the extrapolation of the mass-metallicity relations at low masses, if desired.

\begin{figure}
\centering
\includegraphics[width = 8cm]{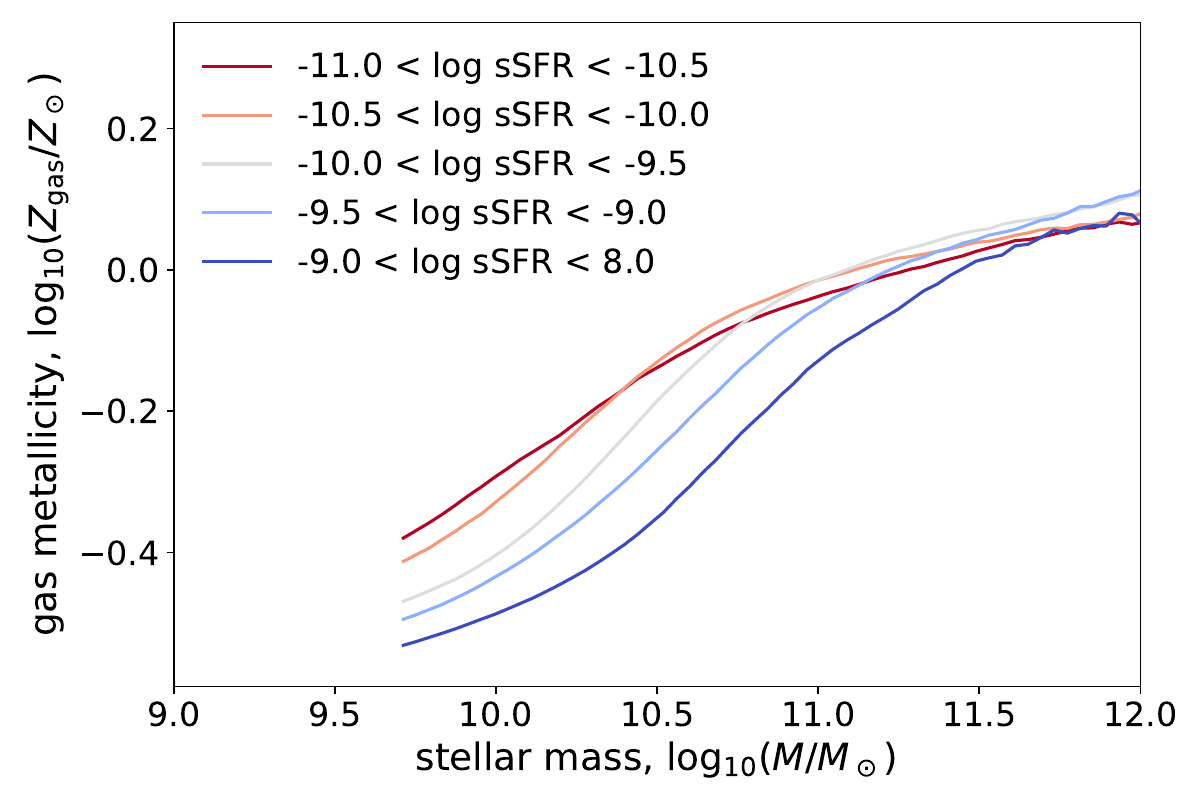}
\caption{Fundamental metallicity relation showing median predicted gas metallicity conditional on stellar mass in bins of $\log10(\mathrm{sSFR})$.}
\label{fig:FMR}
\end{figure}
\subsection{Fundamental metallicity relation}
\label{sec:FMR}
The interplay between star formation and the chemical enrichment of the ISM is expected to result in a relationship between mass, gas-phase metallicity, and star-formation rate -- the so-called fundamental metallicity relation (FMR; \citealp{mannucci2010fundamental, dayal2013physics}).

In Figure \ref{fig:FMR} we show the dependence of the mass-gas metallicity relation with SFR; the second component of the fundamental metallicity relation. We find a clear and smooth negative trend between gas metallicity and SFR for masses up to around $10^{11.5}M_\odot$, with a $0.2-0.3$ dex evolution in the median gas metallicity across the full dynamic range of SFR, across most stellar masses. This is qualitatively consistent with other measurements in the literature \citep{mannucci2010fundamental, dayal2013physics, salim2014, zahid2014fmos, curti2020, thorne2022devils}, and sits roughly in the middle in terms of the magnitude of the trend compared to recent measurements (e.g., \citealp{curti2020} find a trend of up to $0.5$ dex, while \citealp{thorne2022devils} report an overall variation of only $0.13$ dex with SFR).

Whether or not there exists a dependence of the mass-gas metallicity relation with SFR at all (and hence the existence of the FMR as a fundamental plane) is still under debate, with some studies finding a negative trend between gas-metallicity and SFR \citep{mannucci2010fundamental, dayal2013physics, salim2014, zahid2014fmos, curti2020, thorne2022devils}, while others report no significant correlation \citep{sanchez2013mass, sanchez2017mass, sanchez2019sami} or even a positive trend \citep{lara2013galaxy}. Nevertheless, our measurement of the FMR is qualitatively consistent with the most recent measurements \citep{curti2020, thorne2022devils}, and with the physical expectation of a negative trend between gas-metallicity and SFR \citep{mannucci2010fundamental, dayal2013physics}.

We report that inclusion of the gas ionization parameter in our SPS model was essential to recover reasonable inferences about gas-metallicity: without $\mathrm{log}\,U_\mathrm{gas}$, our population-model was unable to recover physically sound predictions for mass-gas metallicity relation and FMR.
\begin{figure*}
\centering
\includegraphics[width = 17.5cm]{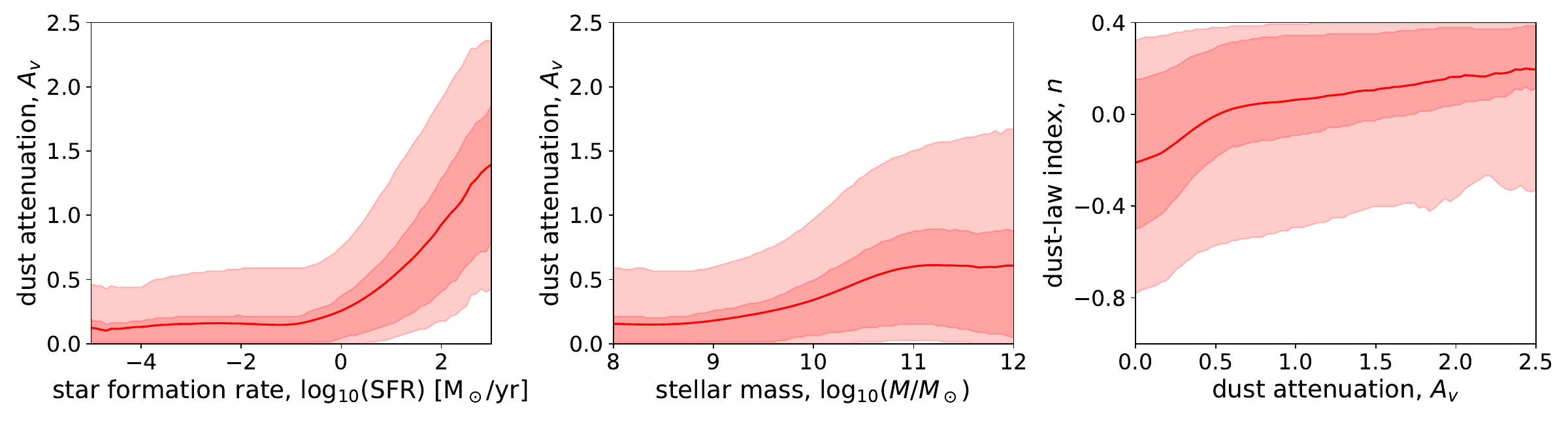}
\caption{Left panel: predicted diffuse dust attenuation as a function of SFR (left) and stellar mass (middle), and the index of the dust attenuation law as a function of dust attenuation (right).}
\label{fig:dust-all}
\end{figure*}
\subsection{Dust attenuation}
\label{sec:dust}
The microscopic properties of dust grains (e.g.\ size, material, etc.) govern their interaction with light, and the direct impact this has on a galaxy's SED (see e.g., \citealp{calzetti2001} or \citealp{draine2003} for a review). Dust grains also impact SEDs through their key role in galaxy star formation, as their surfaces act as favorable media for the formation of molecular hydrogen \citep{gouldsalpeter63, Hollenbach71}. Dust also serves as a key component in regulating heating and cooling, further affecting the star formation cycle \citep{Yamasawa11}. Observations of how dust properties relate to other galaxy characteristics are important in constraining models of galaxy evolution, with key observational targets including the degeneracy between attenuation slope and optical depth, star-dust geometry, and correlations between dust properties with mass and star formation rate \citep[e.g.][]{burgarella2005star, noll09, garn2010, buat2012, zahid2013, kriek2013dust, chevallard2013, reddy2015mosdef, salim2016galex, salmon2016breaking, leja2017, salim2018dust, salim2019, nagaraj2022, lower2022, lower2023}.

Figure \ref{fig:dust-all} (left panel) shows our inferred relationship between (diffuse) dust attenuation and SFR. We see that quiescent galaxies have a tendency toward little or no dust attenuation, although a tail out to non-negligible dust contributions for quiescent galaxies is present. For $\log10(\text{SFR})\gtrsim 0$ the typical level of dust attenuation increases and the spread broadens. Studies of SDSS star-forming galaxies (using H-$\alpha$ emission or the Balmer decrement to measure attenuation, \citealp[e.g.][]{garn2010, zahid2013}) show very similar behavior, with dust attenuation picking up around $\log10(\text{SFR})\gtrsim 0$. More recently, a photometric analysis by \citet{nagaraj2022} using 3D-HST data and the \texttt{Prospector} SPS model also found a strong increase in optical depth at $\log10(\text{SFR})\gtrsim 0$, very similar to our results.

The relationship between dust attenuation and stellar mass (Figure \ref{fig:dust-all}, middle panel) shows a broadening and increase in dust attenuation for galaxies $\gtrsim 10^{10}M_\odot$. \cite{nagaraj2022} also find a similar relationship between dust attenuation and stellar mass, where galaxies $\gtrsim 10^{10}M_\odot$ have higher dust attenuation on average (by around a factor of two) compared to those $\lesssim 10^{10}M_\odot$. Similar results are found by \citet{salim2018dust}, who identify a tendency for higher attenuation values (and a larger scatter) to be seen for more massive galaxies. Our result is also consistent with previous studies of SN Ia host galaxies, where the distribution of extinction values is typically observed to be broader (longer tailed) in galaxies $\gtrsim 10^{10}M_\odot$ \citep[e.g.][]{sullivan2010, childress2013, thorp2021, meldorf2023, grayling2024}. 

In Figure \ref{fig:dust-all} (right panel), we show our inferred relation between dust law index and dust attenuation (for the diffuse dust component). We see a trend towards higher $n$ (shallower attenuation law) for galaxies with higher levels of attenuation, with substantial dispersion ($\sim 0.3$) of the dust index $n$ for any given attenuation $A_V$. This is qualitatively consistent with recent literature \citep{buat2012, kriek2013dust, reddy2015mosdef, salim2018dust, alvarez2019rest, battisti2020strength, nagaraj2022}, and with expectations from radiative transfer calculations \citep[e.g.][]{witt2000, chevallard2013}. We leave an extended quantitative comparison with previous literature to future work.

\subsection{Gas physics}
\label{sec:gas}
\begin{figure}
\centering
\includegraphics[width = 8cm]{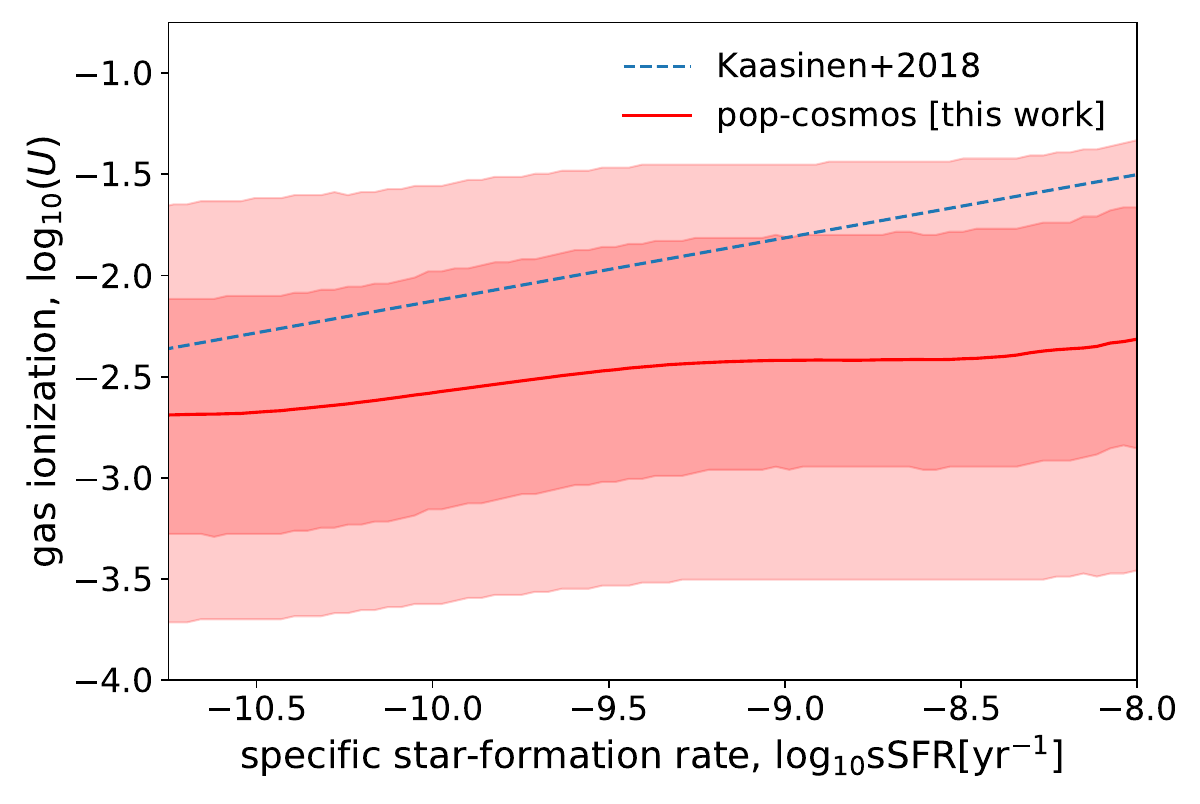}
\caption{Predicted dependence of gas ionization on sSFR. The blue line shows the relation from \citet{kaasinen18}.}
\label{fig:ionization-sSFR}
\end{figure}

While the detailed connection between gas dynamics and star formation is non-trivial, one clear expectation is that gas ionization will increase with increased star formation activity, with massive young stars contributing heavily to the ionizing photon budget. Our population model predicts a clear increasing trend in gas ionization with specific star formation rate (Figure \ref{fig:ionization-sSFR}), qualitatively consistent with previous studies (eg., \citealp{kaasinen18}) and in line with physical expectations. The slope and normalization from \cite{kaasinen18} differ somewhat from our model. We expect relatively weak constraints expected on gas ionization from photometry alone, with emission-lines typically contributing a few percent at most to broad-band fluxes; the level of agreement with \cite{kaasinen18} is very reasonable given the limitations of photometric observations. It is also possible that some differences are due to selection effects in the \cite{kaasinen18} sample.

\subsection{Active galaxies}
\label{sec:AGN}
Figure \ref{fig:triangle} shows some structure in our calibrated population model between AGN and other SPS parameters; most notably, a tendency for the brightest AGN (higher $f_\mathrm{AGN}$) to be redder (higher $\tau_\mathrm{AGN}$), in line with physical expectations (see also Figure \ref{fig:emline_comparisons}). We note that the sharp peak at low values of $f_\mathrm{AGN}$ (and the corresponding spike at intermediate values of $\tau_\mathrm{AGN}$) is an artefact of how we perform the population-model fits, and the information content of the data. For the portion of the galaxy population with little or no AGN contribution, there are no AGN constraints from the data and hence nothing to prefer a sharp peak at some negligible value of $f_\mathrm{AGN}$ over any other distribution over very low values of $f_\mathrm{AGN}$: neither have any discernible impact on the model predictions. Similarly there are no meaningful constraints on $\tau_\mathrm{AGN}$ for galaxies with no AGN contribution; the fact that our model gives all the galaxies with no AGN intermediate values of $\tau_\mathrm{AGN}$ has no impact on our model predictions. While inclusion of AGN is important for population-modeling, drawing detailed inferences about AGN physics likely requires a more sophisticated parameterization of the AGN contribution to the galaxy SEDs.

\subsection{Impact of emission-line calibration on population-level inference}
\label{sec:emline_impact}
A key feature of our model is the ability to self-calibrate emission-line corrections together with the population-model, photometric calibration and uncertainty model. It is informative to explore which population-level inferences are most affected by the emission-line calibrations, and whether those corrections are physically reasonable. Of the all the relations studied in this paper, we find that only the FMR, gas ionization-sSFR relation, and AGN parameters receive any appreciable corrections due to emission-line calibration. This is expected, since the gas metallicity, gas ionization and AGN parameters are expected to be most sensitive to the details of emission-line modeling. Key emission-lines and line-ratios relevant for metallicity and gas-physics (e.g., H$\alpha$, H$\beta$, [OIII] / H$\beta$ and [NII] / H$\alpha$) receive considerable ($\sim$30-60\%) corrections in our model fit (see Table \ref{tab:emlines}).

Figure \ref{fig:emline_comparisons} (Appendix \ref{sec:emline_appendix}) shows model fits with and without allowing the emission-line calibration parameters to vary, for the FMR (top row), gas ionization-sSFR relation (middle row), and AGN luminosity versus optical depth (bottom row). Emission-line calibration induces a $\sim 0.1$dex shift in the normalization of the FMR, with the shape remaining largely unchanged. For the gas ionization-sSFR relation, the model without emission-line calibration barely recovers any correlation between gas ionization and SFR, while the model with emission-line corrections elicits a clearer positive trend between gas ionization and SFR (in line with physical expectations), with around 40\% reduced scatter. For the AGN sector, without emission-line calibration no appreciable constraints on the AGN parameters are recovered, while the model with emission-line corrections recovers a clear (positive) correlation between AGN luminosity and optical depth, in line with physical expectations.

The fact that inclusion of emission-line calibration leads to physically reasonable corrections to the gas-physics and AGN sectors supports the importance of emission-line calibration for obtaining accurate population-level inferences under SPS models. We leave a detailed study of the impact of specific line- and line-ratio corrections and their relation to metallicity, gas- and AGN-physics results to future work.
\section{Discussion}
\label{sec:discussion}
The \texttt{pop-cosmos} population model presented in \S \ref{sec:results} is calibrated down to an $r$-band magnitude of $r<25$. Since selection is corrected for, we expect the model predictions to be valid somewhat deeper than $r<25$, becoming less reliable into the fainter regime where the data is lacking. One of the primary use-cases of our population model in a cosmological inference context is for predicting galaxy redshift distributions from deep, broad-band data from Stage IV surveys such as LSST \citep{alsing2023}. The gold sample for LSST is expected to have a limiting magnitude $r<25.3$, only $0.3$ magnitudes deeper than the COSMOS sample used in this work. Care will need to be taken in examining the extent to which \texttt{pop-cosmos} can extrapolate to $r<25.3$; we leave this to future work. As a stepping-stone to Stage IV cosmological survey applications, we will present \texttt{pop-cosmos} enabled redshift distribution estimation for Stage III (KiDS) data in a companion paper (Loureiro et. al., in prep.), as well as the utility of \texttt{pop-cosmos} as an improved model and prior for individual galaxy photo-$z$ estimation \citep{thorp24_mcmc}.

In the context of calibrating an accurate galaxy population model for improving cosmological analyses (e.g., \citealp{alsing2023, moser2024}), the galaxy-evolution results presented in \S \ref{sec:results} are a essentially side-effect of pursuing accurate redshifts. Nonetheless, the advancement in methodology means that many of the inferred scaling relations may be better measured by our new framework. 

Measurement of the FMR, mass-metallicity-redshift and sSFR-gas ionization relations has generally been considered challenging or impossible from photometric data alone. Nonetheless, in \S \ref{sec:mass-met}-\ref{sec:gas} we present measurements of these relations that are qualitatively consistent with previous results (from spectroscopic data). This opens up an exciting new approach to measuring these quantities, which merits further investigation\footnote{See also \cite{thorne2022devils} for a recent measurement of the mass-metallicity-redshift and FMR relations from photometric data.}. Even where our predictions about gas-phase physics do not agree in detail with spectroscopic measurements (likely due to the limitations of photometric data), we do not expect this to have a significant effect on the photometric redshift program: corrections to broad-band photometry should be tiny\footnote{Second-order gas-physics parameters (i.e., gas-metallicity and ionization) will have a $\lesssim 2-3\%$ effect on broad-band fluxes in the vast majority of cases, with corrections to their population-level distributions and correlations with other parameters being even less significant.}.

In \S \ref{sec:data-space} we saw that our model predictions for galaxy colors are accurate to within observed calibration biases between different photometric extraction methods (e.g., the \texttt{Farmer} vs. \texttt{Classic} versions of the COSMOS photometry; \citealp{weaver2023}). There is hence no evidence that additional complexity in the SPS model is justified at this stage to improve the population-model predictions. Nevertheless, the scalability of our simulation-based inference approach opens up the possibility of including further extensions (and parameters) within the SPS model, while remaining computationally feasible.

We established that self-calibration of emission-line corrections was important for drawing reasonable population-level predictions of the gas- and AGN-sector parameters, and was shown to improve photometric redshift inferences in the COSMOS data (see, e.g., \citealt{Alarcon_2021, leistedt2023hierarchical}). While parameterizing the mean bias in the most important emission-lines captures the leading order correction, in practice emission line strengths will be a strong function of the SPS parameters. It would be straightforward to incorporate parameter-dependent emission-line corrections in our calibration model; we leave this to future work.

Recently, another forward modeling-based approach to inferring the population distribution of galaxy parameters has been presented(\textsc{popsed}; \citealp{li2024popsed}), also making use of the OT distance as an objective function in their inference procedure. They use normalizing flows as their population distribution over SPS parameters, with a 12-parameter SPS model following \citet{hahn2023provabgs}, and an SPS emulation scheme similar to \citet{alsing2020}. They demonstrate the method on broad-band SDSS $ugriz$ photometry for a sample of galaxies from the Galaxies and Mass Assembly survey (GAMA; \citealp{driver2011, baldry2018}), with a depth of $r<19.8$ and at relatively low redshift $z\lesssim0.45$, showing in particular that they can recover the star-forming main sequence (c.f.\ our \S\ref{sec:SFS} and Figure \ref{fig:SFS}).

Our work goes further than \cite{li2024popsed} by constructing and fitting a comprehensive forward model for the data, including: a flexible population-model; state-of-the-art (16-parameter) SED model; self-calibration of the data-modeling (noise and calibration); and explicit treatment of selection. By utilizing a larger, deeper galaxy sample, we cover the depth ($r<25$) and redshift range ($z\lesssim 3.5$) required for modeling Stage III and IV wide-deep galaxy surveys. The broad wavelength range covered by the $26$-bands used here (including intermediate and narrow bands) also allows us to constrain a comprehensive range of galaxy evolution physics, for a diverse galaxy sample. As a result our calibrated population-model can faithfully predict galaxy colors over a wide range of wavelengths and redshifts, with direct utility in a cosmological inference context for Stage III and IV surveys.

Another forward modeling-based approach has been developed by \citet{moser2024} and applied to image-level data from HSC \citep{aihara2022}. Their population model is parametric, with galaxy spectra built up from \textsc{Kcorrect} templates \citep{blanton2017}, and source detection within their simulated images being handled by \textsc{SExtractor} \citep{bertin1996}. Their inference is carried out using an approximate Bayesian computation (ABC) scheme (also employed by \citealp{tortorelli2020, tortorelli2021}). Our work differs from theirs in utilizing a continuous SPS model (rather than templates) for galaxy SEDs, and a flexible (diffusion-model) parameterization of the population-model, while jointly calibrating the population- and data-model simultaneously. OT optimization is also expected to scale favourably to high-dimensional problems on large datasets, where ABC quickly becomes computationally infeasible due to the high number of simulations required (see e.g., \citealp{alsing2019}). Nevertheless, forward modeling at the level of images represents an important advance, and may be necessary for including and correcting for image-based selection cuts in future analyses.
\section{Conclusions}
\label{sec:conclusions}
We have presented \texttt{pop-cosmos}: a comprehensive population model fit to a large, deep, flux-limited sample of galaxies from COSMOS. We constructed a detailed forward model for the COSMOS data, including a flexible diffusion-model parameterzation of the population-distribution of galaxy characteristics, a state-of-the-art (16-parameter) SPS model, and a detailed data-model describing the observation, calibration and selection processes. By comparing synthetic and real data in a simulation-based inference setting, we were able to jointly fit the population-model while self-calibrating the data- and calibration-model parameters in a self-consistent fashion. As a result, we obtained a robustly calibrated population model describing galaxies down to $r<25$ and out to redshift $z\simeq 3.5$. 

Our population model is able to faithfully reproduce galaxy colors (to within the limitations of the photometric calibration of the COSMOS data), and encodes a comprehensive and compelling picture of galaxy evolution processes. This represents the first time that it has been possible to jointly infer the full, complex web of dependencies between galaxy characteristics, together with the photometric noise, data- and model-calibration, and principled correction of selection: a key milestone in the analysis of large galaxy surveys. 

Accurate galaxy population models calibrated to large, deep, narrow band (or spectroscopic) data are of key importance in drawing robust cosmological measurements from galaxy surveys. We expect the \texttt{pop-cosmos} model and its successors to open up new capabilities in accurate redshift estimation from photometric data, eliminating systematics in transient cosmology due to correlations between host galaxy properties and supernovae, and in modeling and inferring the galaxy-halo connection.

\acknowledgments
\textbf{Author contributions.} 
We outline the different contributions below using keywords based on the CRediT (Contribution Roles Taxonomy) system.\\
{\bf JA:} conceptualization, methodology, software, validation, formal analysis, investigation, data curation, writing - original draft, writing - editing \& review, visualization. 
{\bf ST:} methodology, validation, visualization, formal analysis, writing - editing \& review.
{\bf SD:} validation, visualization, formal analysis, writing - editing \& review.
{\bf HVP:} conceptualization, methodology, validation, visualization, writing - editing \& review, supervision, funding acquisition. 
{\bf BL:} data curation, formal analysis, validation, writing - editing \& review.
{\bf DM:}  methodology, validation, writing - editing \& review, funding acquisition. 
{\bf JL:} validation, writing - editing \& review.

\textbf{Data availability.}
The COSMOS2020 catalog\footnote{\url{https://cosmos2020.calet.org}} \citep{weaver2020} and pre-processing scripts\footnote{\url{https://github.com/cosmic-dawn/cosmos2020-readcat}} are publicly available. Other data underlying this article will be shared on reasonable request to the authors.

\textbf{Acknowledgements.} We thank George Efstathiou and Arthur Loureiro for useful discussions. JA, ST, SD and HP have been supported by funding from the European Research Council (ERC) under the European Union's Horizon 2020 research and innovation programmes (grant agreement no. 101018897 CosmicExplorer). This work has been enabled by support from the research project grant ‘Understanding the Dynamic Universe’ funded by the Knut and Alice Wallenberg Foundation under Dnr KAW 2018.0067. HVP was additionally supported by the G\"{o}ran Gustafsson Foundation for Research in Natural Sciences and Medicine. HVP and DM acknowledge the hospitality of the Aspen Center for Physics, which is supported by National Science Foundation grant PHY-1607611. The participation of HVP and DM at the Aspen Center for Physics was supported by the Simons Foundation. This research also utilized the Sunrise HPC facility supported by the Technical Division at the Department of Physics, Stockholm University.  BL is supported by the
Royal Society through a University Research Fellowship. This study utilizes observations collected at the European Southern Observatory under ESO programme ID 179.A-2005 and 198.A-2003 and on data products produced by CALET and the Cambridge Astronomy Survey Unit on behalf of the UltraVISTA consortium. 

\appendix

\section{Comparison of population-level inferences with and without emission-line calibration}
\label{sec:emline_appendix}
\begin{figure*}
\centering
\includegraphics[width = \linewidth]{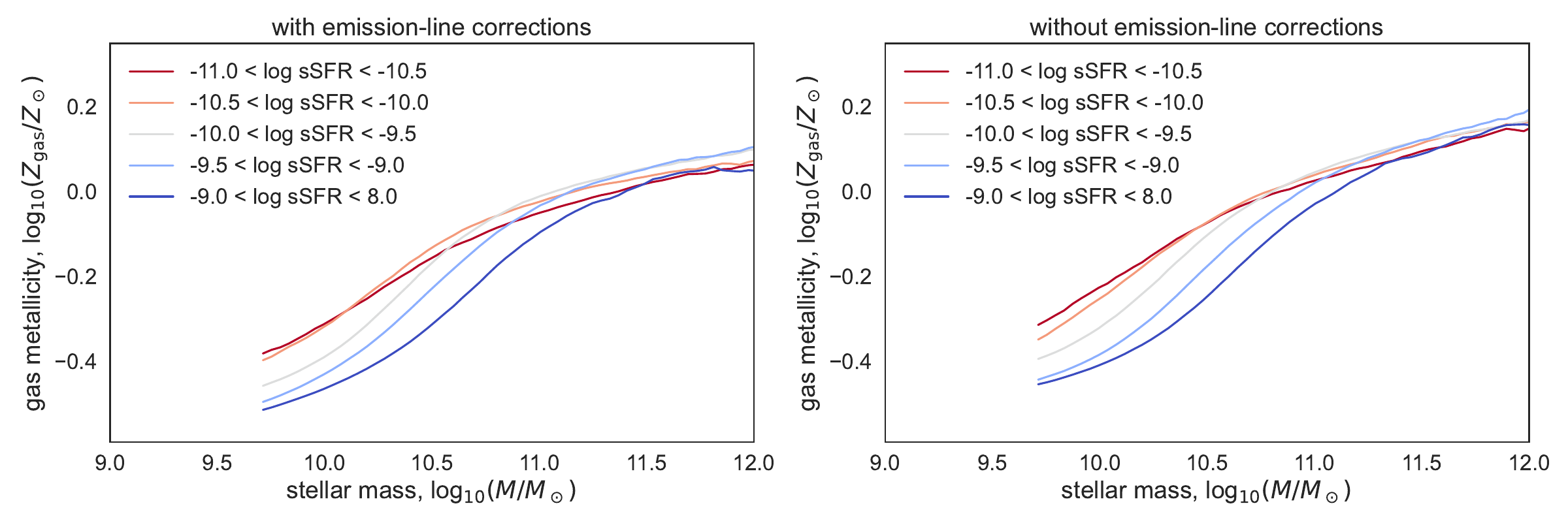}
\includegraphics[width = \linewidth]{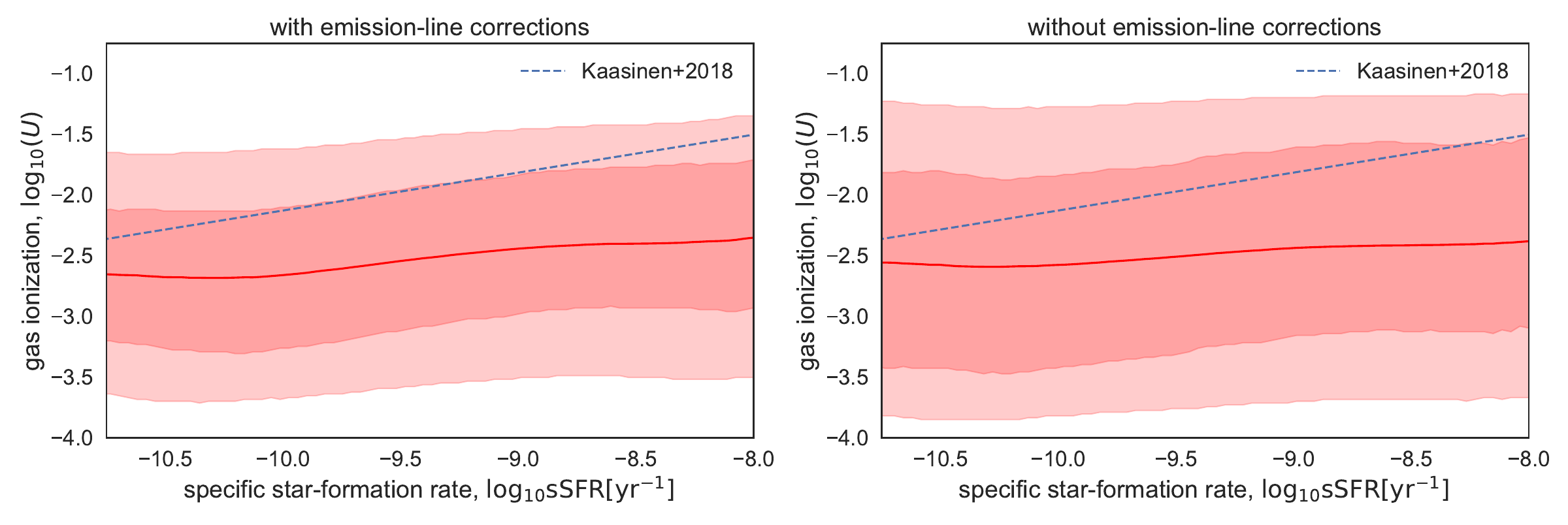}
\includegraphics[width = \linewidth]{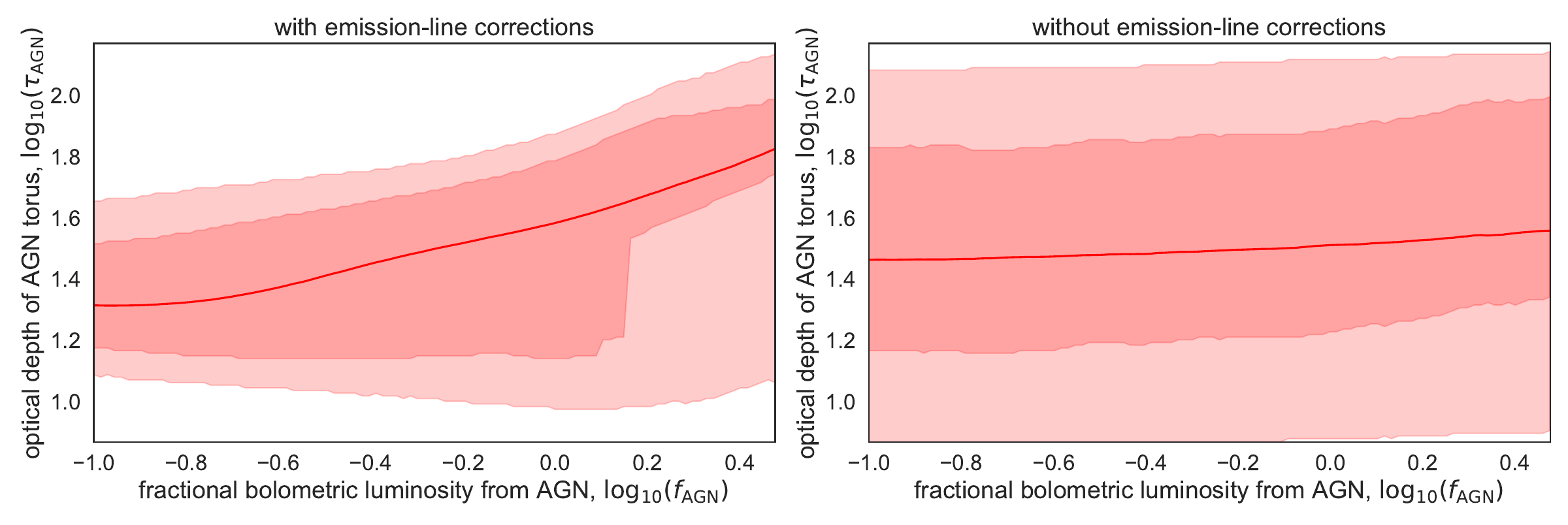}
\caption{Comparison of \texttt{pop-cosmos} fits with (left column) and without (right column) emission-line calibration, for three population-level quantities most impacted by emission-line calibration: the fundamental-metallicity relation (FMR; top row), gas ionization-sSFR relation (middle row), and the relationship between the AGN luminosity and optical depth (bottom row). }
\label{fig:emline_comparisons}
\end{figure*}

Figure \ref{fig:emline_comparisons} shows side-by-side comparisons of \texttt{pop-cosmos} fits with and without emission-lines for the three population-level quantities that are most sensitive to emission-line modeling. We see that the FMR (top row) gets a $\sim 0.1$dex correction in normalization due to emission-line calibration, while the shape of the FMR is broadly unchanged. For the gas ionization-sSFR relation (middle row), without emission-line calibration the model find little or no appreciable connection between gas ionization and star formation rate. With emission-line calibration included, a more significant positive relation between gas ionization and SFR emerges (in line with physical expectations) with around 40\% less scatter, albeit still shallower than the relation from \citet{kaasinen18} (calibrated to spectra). For the AGN sector (bottom row), without emission-line calibration the model is unable to recover any appreciable constraints on the AGN parameters, while the model with emission-line corrections recovers a clear correlation between AGN luminosity and optical depth, in line with physical expectations. 
\clearpage
\bibliography{spsbhm}

\end{document}